\documentclass[%
 reprint,
 amsmath,amssymb,
 prb,
]{revtex4-1}

\usepackage{graphicx}
\usepackage{dcolumn}
\usepackage{bm}
\usepackage{url} 
\begin{document}

\preprint{APS/123-QED}

\title{Band alignment and scattering considerations for enhancing the thermoelectric power factor of complex materials: The case of Co-based half-Heuslers}

\author{Chathurangi Kumarasinghe}
\email{*Chathu.Kumarasinghe@warwick.ac.uk} 
\author{Neophytos Neophytou}
\affiliation{School of Engineering, University of Warwick, Coventry, CV4 7AL, UK}


\date{\today}

\begin{abstract}
Producing high band and valley degeneracy through aligning of conducting electronic bands is an effective strategy to improve the thermoelectric performance of complex bandstructure materials. Half-Heuslers, an emerging thermoelectric material group, has complex bandstructures with multiple bands that can be aligned through band engineering approaches, giving us an opportunity to improve their power factor. Theoretical calculations to identify the outcome of band engineering usually employ detailed density functional theory for bandstructure calculations, but the transport calculations are kept simplistic using the constant relaxation time approximation due to the complications involved with detailed scattering physics. In this work, going beyond the constant relaxation time approximation, we perform an investigation of the benefits of band alignment in improving the thermoelectric power factor under different density of states dependent scattering scenarios. As a test case we consider the Co-based \textit{p}-type half-Heuslers TiCoSb, NbCoSn and ZrCoSb. First, using simplified effective mass models combined with  Boltzmann transport, we investigate the conditions of  band alignment that are beneficial to the thermoelectric power factor under three different carrier scattering scenarios: i) the usual constant relaxation time approximation, ii) intra-band scattering restricted to the current valley with the scattering rates proportional to the density of states as dictated by Fermi's Golden Rule, and iii) both intra- and inter-band scattering across all available valleys, with the rates determined by the total density of states at the relevant energies. We demonstrate that the band-alignment outcome differs significantly depending on the scattering details. Next, using the density functional theory calculated bandstructures of the half-Heuslers we study their power factor behavior under strain induced band alignment. We show that strain can improve the power factor of half-Heuslers, but the outcome heavily depends on the curvatures of the bands involved, the specifics of the carrier scattering mechanisms and the initial band separation. Importantly, we also demonstrate that band alignment is not always beneficial to the power factor. In addition, we show that the bandstructure itself can undergo changes as the bands are aligned in practice, which further affect the band alignment optimization. Our work illustrates the importance of going beyond the constant relaxation time approximation, as well as understanding how the bandstructure of each material behaves when considering band alignment.      
\end{abstract}

\pacs{Valid PACS appear here}
\maketitle


\section{\label{sec:level10}Introduction }
Thermoelectric (TE) materials are capable of directly converting heat into electricity and vice versa, and are useful in power generation from waste heat \cite{beretta2018thermoelectrics,bell2008cooling,shakouri2011recent,koumoto2013thermoelectric,zhang2015thermoelectric}. The ability of a material to produce thermoelectric power efficiently is quantified by the dimensionless figure of merit: 
\begin{equation}\label{Eq1}
ZT=\sigma S^2 T/\kappa,
\end{equation}
where $\sigma$ is the electrical conductivity, $S$ is the Seebeck coefficient, $T$ is the temperature and $\kappa$  is the thermal conductivity of the material. For a high $ZT$, a high electrical conductivity, a high Seebeck coefficient, i.e., a high power factor ($ \sigma S^2$), and a low thermal conductivity $\kappa$ are desirable. However, simultaneous optimization of these parameters remains a challenge due to their complicated adverse interdependencies. Recently emerged advanced thermoelectric materials, such as half-Heusler alloys\cite{yang2008evaluation, xi2016band, huang2016recent, page2016first, xie2012recent,bos2014half,casper2012half,chen2013recent}, SnSe, PbTe, and BiTe based compounds \cite{heremans2008enhancement,zhao2014ultralow,wolfing2001high}, clathrates \cite{kleinke2009new, nolas1998semiconducting}, skutterudites \cite{sales1996filled, nolas1999skutterudites}, to name a few, possess complex crystalline and electronic bandstructures, exhibiting multiple anisotropic valleys with high degeneracies capable of contributing to conduction. Such features can be useful to overcome the unfavorable interdependencies at least of the conductivity and the Seebeck coefficient by application of bandstructure engineering approaches to improve the power factor\cite{snyder2011complex,zhao2013all,tan2014high}.

One of the most promising and commonly employed bandstructure engineering approaches in multi-band materials is to increase the valley or orbital degeneracy near conduction or valence bands edges \cite{pei2011convergence, tang2015convergence, norouzzadeh2016classification}, referred to as ‘band convergence’ or ‘band alignment’. The idea is that when multiple bands contribute to transport, the conductivity, and therefore the power factor, will improve. In bulk materials, bandstructures can be manipulated by applying strain, doping, alloying,  and second phasing with other suitable structures \cite{li2015enhancement, tan2016rationally, imasato2018band,liu2012convergence, thompson2006future}. At nanoscale, additional means of influence such as modifying the size, shape and the chemical surrounding, to name a few, are available \cite{leu2008ab, yan2008simple, neophytou2010large, neophytou2010analysis}. 

In addition to good thermoelectric performance, the ideal thermoelectric material should have low toxicity, relatively inexpensive elemental composition, good thermal stability, and be easily produced on a large scale.  Half-Heusler alloys are one of the few classes of materials that fulfill the above requirements \cite{yang2008evaluation, misra2014enhanced}. They are known to have impressive power factors, but unfortunately high thermal conductivities \cite{zhou2018large, misra2014enhanced, fu2014high, huang2016recent}. As a consequence, much work on half-Heuslers focus on lowering of the thermal conductivity by introducing multi-scale defects, manipulating grain sizes, and alloying with elements of large mass contrast. High $ZT$ values close to 1.5 have been achieved under moderate temperatures using such techniques \cite{fu2015realizing, graf2011simple}. Their complex electronic structure, however, provides opportunities to further optimize the inherently good power factors through bandstructure engineering.

The majority of theoretical work related to bandstructure engineering (as well as material screening), involves calculating the bandstructures using \textit{ab initio} density functional theory (DFT), which is then used in conjunction with the Boltzmann Transport Equation (BTE) in the relaxation time approximation to compute the thermoelectric coefficients \cite{scheidemantel2003transport,yang2008evaluation}. Due to the complexities in accurate scattering treatment and the variety of scattering mechanisms, it is common to adopt a constant relaxation time ($\tau$) approximation (usually  $\tau \approx 10^{-14}$ s is used at 300\thinspace K \cite{berche2018fully, scheidemantel2003transport}). However, it is quite evident that such a simplification will fail and lead to false estimations of the power factor, particularly when multiple bands participate in transport, such as in studies of band-alignment optimization. Simply, while aligning the bands can increase the number of carriers available for conduction, from simple Fermi's Golden Rule considerations, it can also increase the number of states that carriers scatter into, which hinders the carrier transport. Therefore, the energy dependence of the scattering mechanisms, as well as the specifics of intra- or inter-valley scattering considerations are important in identifying if a given bandstructure engineering approach leads to an improved power factor, or not\cite{lundstrom2009fundamentals,pshenay2010effect,witkoske2017thermoelectric}.

In this work, we investigate the role of band alignment in improving the power factor in complex bandstructure materials, by considering three possible scattering conditions: i) the commonly employed constant relaxation time ($\tau_\mathrm{C}$), ii) scattering proportional to the density of states of the band, but restricting to only intra-valley scattering ($\tau_\mathrm{IV}(E)$), and iii) scattering proportional to the total density of states, allowing both intra- and inter-valley  and inter- and intra-band scattering ($\tau_\mathrm{IIV}(E)$). We note that one needs to understand the influence of all three scenarios, as to date, there is almost complete lack of understanding, either theoretical or experimental, in providing evidence in the true nature of scattering in these materials. 
 
 As a test case, we use the bandstructures of the Co-based \textit{p}-type half-Heuslers, TiCoSb, NbCoSn, ZrCoSb and ZrCoBi, which have multiple valleys (or carrier pockets) with multiple bands, that can be aligned at the valence band edge (VB$_0$). We show that when attempting to improve the power factor of materials through band alignment, depending on i) the scattering considerations, ii) the masses of the aligned bands, iii) initial band separation and iv) the changes that appear in the bandstructure upon alignment, different outcomes to the power factor are reached. We show that contrary to current view, band alignment is not always beneficial to the power factor, in fact, in some cases it is misalignment that leads to improvements. We then present in detail the conditions for power factor improvements through simplified equations that would prove useful to materials scientists. We further consider the use of strain as a band alignment strategy for these half-Heuslers for power factor improvements. We show that strain can indeed align the bands of Heusler materials, and this can result in even up to a 40\% improvement in the power factor.  

The paper is organized as follows:  In Section II we describe our theoretical approach.  In Section III, we start with two simple parabolic bands to illustrate optimum conditions for band alignment under the three different carrier scattering considerations. In Section IV we first describe simplified, computationally inexpensive non-parabolic effective mass models derived out of DFT calculated bands to identify potential improvements in the power factor of Co-based half-Heuslers as a result of band alignment.  Then, in section V, using more computationally expensive DFT and semi-classical Boltzmann transport calculations, we investigate how strain can be used in reality to achieve band alignment. Finally, in Section VI we conclude.

\section{\label{sec:level16}Methods }

\begin{figure*}
\includegraphics[width=0.95\textwidth]{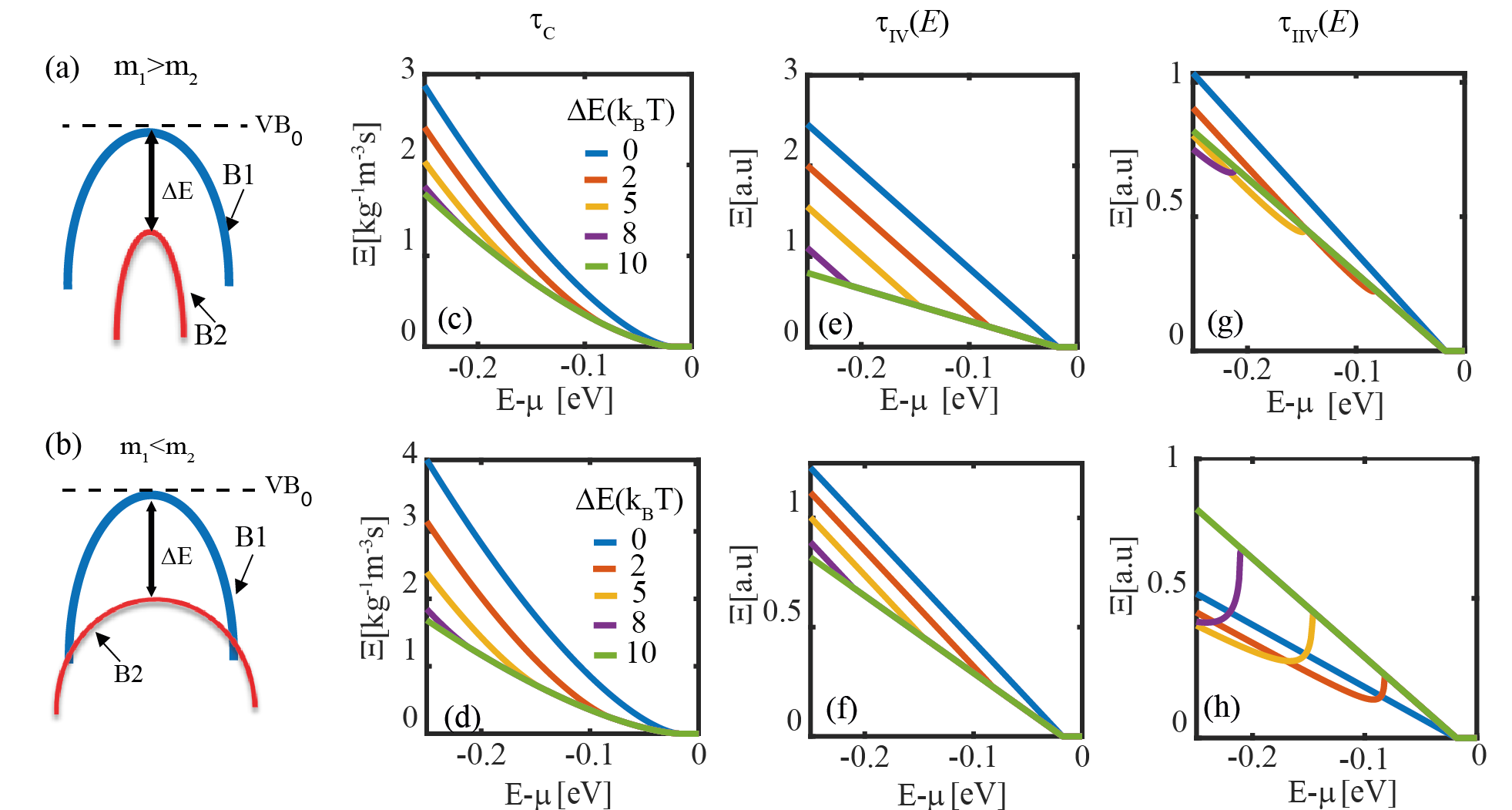}
\caption{\label{fig1} Transport distribution (TD) functions for different band mass combinations (row-wise) under different scattering scenarios (column-wise). Band B1, which is already aligned with the valence band edge (VB$_0$) has a mass $m_1$ and band B2, which is below it has a mass $m_2$. (a) Schematic representation of aligning a lighter band ($m_1=1m_0$, $m_2=0.5m_0$) in the first row, and (b) aligning a heavier band ($m_1=1m_0$, $m_2=2m_0$) in the second row. The displacement between the two bands is given by $\Delta E$. The first row shows the TD functions for aligning a lighter band in case of (c) constant rate of scattering ($\tau_\mathrm{C}$), (e) intra-band scattering only ($\tau_\mathrm{IV}(E)$), (g) and inter- and intra-band scattering ($\tau_\mathrm{IIV}(E)$). The second row shows the TD functions for aligning a heavier band in case of (d) constant rate of scattering ($\tau_\mathrm{C}$), (f) intra-band scattering only ($\tau_\mathrm{IV}(E)$),  and (h) inter- and intra-band scattering ($\tau_\mathrm{IIV}(E)$).  }
\end{figure*} 

\subsection{\label{sec:level13}Boltzmann Transport Equation}
To compute the thermoelectric coefficients we employ the Boltzmann transport formalism within the relaxation time approximation (RTA). The materials under consideration are described using effective mass approximations (parabolic and non-parabolic), as well as DFT extracted bandstructures. The thermoelectric coefficient tensors, electrical conductivity $\sigma_{\alpha,\beta} (T,E_F )$ and Seebeck coefficient $S_{\alpha,\beta} (T,E_F )$, are expressed as \cite{mahan1996best, neophytou2011effects}: 
\begin{equation}\label{Eq2}
    \sigma_{\alpha\beta}\left(T,E_F\right)=e^2 \int \mathrm{\Xi}_{\alpha\beta}\left(E\right)\Big(-\frac{\partial f(E,E_F,T)}{\partial E}\Big) dE
\end{equation}

\begin{equation}\label{Eq3}
  S_{\alpha\beta}\left(T,E_F\right)= \frac{e\int \mathrm{\Xi}_{\alpha\beta}\left(E\right)(E-E_F)(-\frac{\partial f(E,E_F,T)}{\partial E})dE}{ T\sigma_{\alpha\beta}\left(T,E_F\right)} 
\end{equation}
where $e$  is the charge of an electron and $f(E,E_F,T)$  is the Fermi distribution function at a given temperature $T$ and a chemical potential $E_F$. The transport distribution (TD) function, $\Xi_{\alpha\beta}\left(E\right)$ which is a function of carrier energy $E$, is given by:
\begin{equation}
\Xi_{\alpha\beta}(E)=\sum_{i,\mathbf{k}}\tau_{i,\mathbf{k}}(E)v_\alpha(i,\mathbf{k})v_\beta (i,\mathbf{k}) \delta(E-E_{i,\mathbf{k}})
\end{equation}
where $i$ and $\mathbf{k}$  are the band index and the k-point, respectively.  The electron relaxation time is denoted by $\tau_{i,\mathbf{k}}(E)$ and $v_\alpha(i,\mathbf{k})\ (\alpha={x,y,z})$ represents the $\alpha^{th}$ component of the group velocity $ \mathbf{v}(i,\mathbf{k})$ which can be derived from the slope of the bands in the bandstructure as: 
\begin{equation}
    v(i,\mathbf{k})= \frac{1}{\hbar}\mathrm{\nabla}_\mathbf{k} E_{i,\mathbf{k}}. 
\end{equation}

It can be seen from Eqs.\ref{Eq2} and \ref{Eq3} that the conductivity will increase with the TD function, while the Seebeck coefficient has a more complicated relation, with the contribution of higher energies being weighted more. Therefore, the Seebeck coefficient depends on the energy derivative of the TD function. Numerical calculations for the transport coefficients were carried out using the BoltzTraP code\cite{madsen2006boltztrap} (in the case of constant RTA and DFT calculated bandstructures), in combination with our own developed codes (for the cases of energy-dependent RTA and parabolic/non-parabolic effective mass approximation bandstructures), as noted below in each described case.

\subsection{\label{sec:level17}\textit{Ab initio} electronic structure calculations}

We have performed \textit{ab initio} DFT calculations for the Co-based half-Heuslers, NbCoSn, TiCoSb, ZrCoSb and ZrCoBi with the QUANTUM ESPRESSO package \cite{giannozzi2009quantum}. Projector augmented wave technique was used with the PBE-GGA functional and a kinetic energy cutoff greater than 60\thinspace Ry was used for the wavefunctions. An energy convergence criterion of $10^{-8}$\thinspace Ry for self-consistency was adopted throughout our calculations. For transport property calculations, a 15x15x15 Monkhorst–Pack k-point sampling was used for the primitive unit cell with three atoms. Calculations using denser k-points were also carried out to confirm the convergence of the results. We calculate an average of transport quantities in x, y and z crystalline directions, when using DFT derived bands in combination with BoltzTraP. Using the information of three crystallographic orientations, rather than the full anisotropy is known to give sufficient accuracy in thermoelectric calculations \cite{zahedifar2018band,zhao2016ultrahigh,singh2010doping}. In the DFT calculations, spin orbit coupling (SOC) effects were not considered. SOC introduces band splitting and changes in the separation between the different valleys, but these effects do not affect our analysis or our qualitative conclusions. In fact, SOC effects are insignificant in the TiCoSn and ZrCoSb cases, whereas in the case of NbCoSn, SOC effects the upper valleys slightly \cite{zahedifar2018band,zhu2018discovery}. For details of bandstructure comparisons with and without SOC see our calculations in  Supplemental Matrial \cite{supplimentaryInfo}. 

\subsection{\label{sec:level14}Parabolic band approximation}

For our initial, first-order understanding of the effect of band alignment, we construct a bandstructure consisting of two parabolic bands ($E= \hbar^2\mathbf{k}^{2}/(2m)$) with different effective masses, $m$ for each band (as shown in Figs.\ref{fig1}(a) and \ref{fig1}(b)). We then examine the thermoelectric power factor upon aligning these bands under different scattering RTA scenarios: i) the commonly employed constant relaxation time, $\tau_i(E)=10^{-14}$\thinspace s, ii) scattering proportional to the density of states of the band, but restricting to only intra-valley scattering ($\tau_i(E)\propto 1/\mathrm{DOS}_i(E)$), and iii) scattering proportional to total density of states, allowing both intra- and inter-valley  and inter- and intra-band scattering  ($\tau_i(E)\propto 1/\sum_{i}\mathrm{DOS}_i(E)$). We have not considered inter- and intra-band scattering with only intra -valley scattering scenario because above mentioned scenarios are sufficient to provide a general understanding of the effect of scattering. Since under the parabolic band approximation, the velocity and density of states of each band is $v_i(E)= {(2E/m_i)}^{1/2}$ and ${\rm DOS}_i\left(E\right)={2^{1/2}m}_i^{3/2}N_iE^{1/2}/{(\pi}^2\hbar^3)$, respectively, the TD function for valence bands is reduced to:

\begin{equation}\label{Eq6}
   \mathrm{\Xi}\left(E\right)\propto \sum_{i} N_i \tau_i(E) { m}_i^{\frac{1}{2}} E^\frac{3}{2}  H\left(-\Delta E_i\right)
\end{equation}
\\
where the subscript $i$ indicates each band, $\Delta E_i$ indicates the distance to the band edge from the valence band edge, $N_i$ indicates the band degeneracy and $H\left(\Delta E_i\right)$ indicates the Heaviside step function. For conduction bands, $H\left(-\Delta E_i\right)$ should be replaced by $H\left(\Delta E_i\right)$. We label the band that is already at the valence band edge VB$_0$ as B1, and the second, as the \lq{aligning band}\rq \thinspace B2. 

We note that which of the three scattering scenarios is the most appropriate, is not possible at this point to determine – it might be that it will be different for different materials, different energies in the same material, or even a combination of all three in the same material. Experimental studies could provide guidance towards understanding the nature of scattering in half-Heuslers, however, data are sparse at the moment, and mostly for alloys and for specific charge densities. Even when it comes to the constant relaxation times, the actual values can differ by orders of magnitude. Indeed, in the Supplementary Information\cite{supplimentaryInfo} we analyzed data from 2 experiments for doped-TiCoSb alloys \cite{wu2009effects,wu2007thermoelectric}, which point out to larger relaxation times, however, we still use below the more commonly employed $\tau =10^{-14}$~s. We have also performed ab-initio electron-phonon scattering calculations using the EPW package \cite{ponce2016epw}, for TiCoSb and ZrCoSb in an attempt to understand the nature of scattering, where it seems that the relaxation times follow roughly the downward trend of $1/\mathrm{DOS}(E)$ for TiCoSb, whereas for ZrCoSb a more constant trend followed by a rough $1/\mathrm{DOS}(E)$ at high energies, but extracting further details seems at this point is difficult (see Supplementary Information \cite{supplimentaryInfo}). Thus, separating the three cases and studying them individually, provides a first order understanding on the effect of each scattering scenario. 

We perform band alignment investigations for the three different scattering rates for two scenarios: 1) band B1 has a heavier mass ($m_1=1m_0$)  than the aligning band B2 $\left(m_2=0.5m_0\right)$, and 2) B1 has a lighter mass ($m_1=1m_0$ )  compared to the aligning band B2  ($m_2=2m_0$) as shown in Fig.\ref{fig1}(a) and Fig.\ref{fig1}(b), respectively. The value $m_0$ is the rest mass of the electron. We assume the band degeneracy, $N=1$, for this study. Band B1 is already aligned with VB$_0$, and we then bring band B2 gradually closer to B1 by reducing $\Delta E$. We first examine the TD functions to understand the trends of our results, because the influence of the different bands at different $\Delta E$  appears there clearly. Figure \ref{fig1}(c)-(h) shows the TD functions for two the bands with masses $m_1$ and $m_2$ for different levels of alignment in units of $k_\mathrm{B}T$, varying between unaligned ($\Delta E=10\thinspace k_\mathrm{B}T$) to fully aligned ($\Delta E=0$). Column-wise we show results for the three different scattering situations we have considered, as labeled. Row-wise we show results in the case where we bring a lighter/heavier band into transport, respectively. We note that all our calculations are performed at $T = 300$\thinspace K.  

\begin{figure*}
\centering
\includegraphics[width=0.95\textwidth]{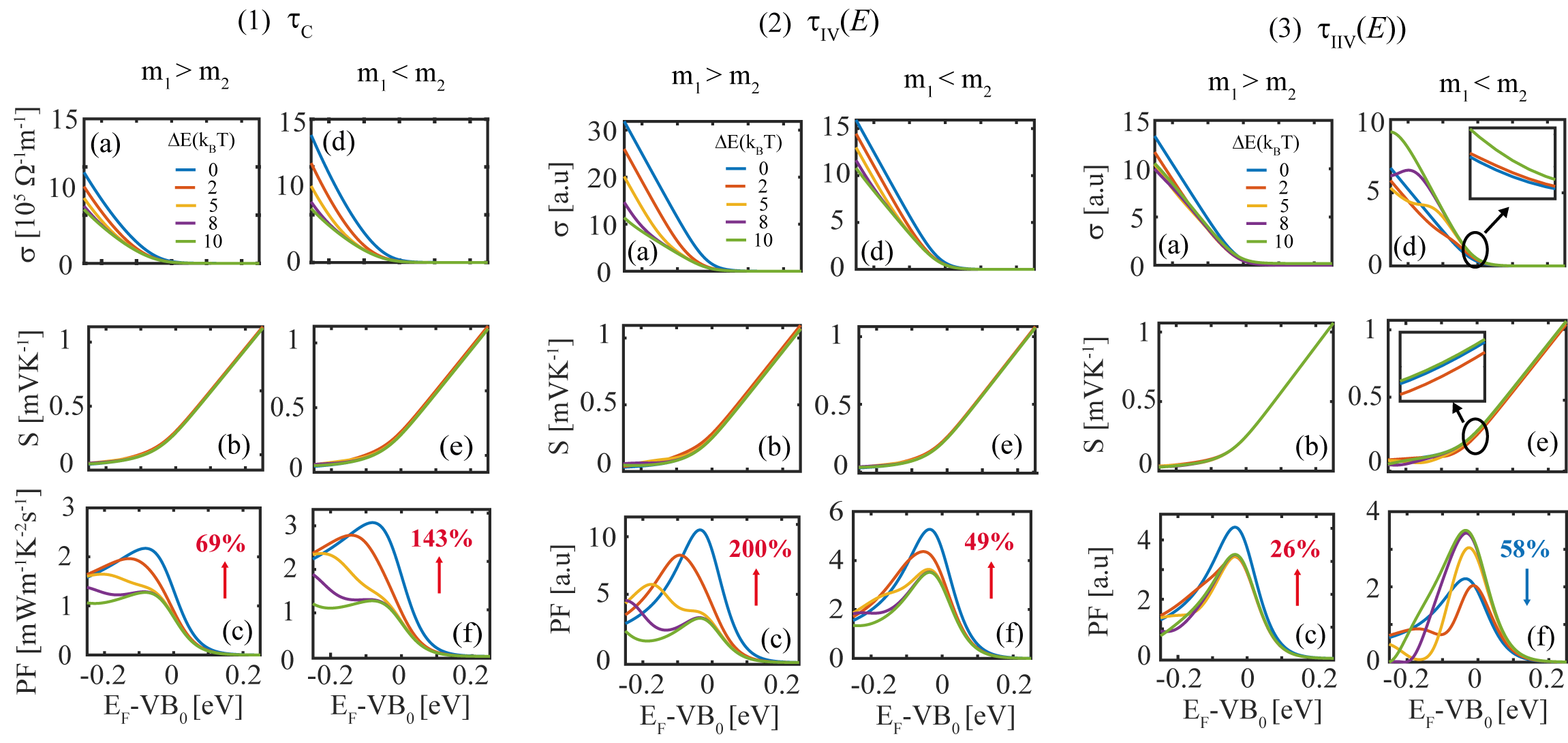}
\caption{\label{fig2} Thermoelectric coefficients when aligning bands with different mass combinations under three scattering scenarios: Panel (1) for constant rate/time of scattering ($\tau_\mathrm{C}$), panel (2) for intra-band scattering only ($\tau_\mathrm{IV}(E)$) and panel (3) for inter- and intra-band scattering ($\tau_\mathrm{IIV}(E)$). Each panel shows the coefficients: (a) electronic conductivity, (b) Seebeck coefficient, and (c) power factor when the mass of B1 is larger than that of B2, i.e. $m_1>m_2$ ($m_1=1m_0$, $m_2=0.5m_0$). Sub-figures (d-f) in each panel show the (d) electronic conductivity, (e) Seebeck coefficient, and (f) power factor when the mass of B1 is smaller than that of B2, i.e. $m_1<m_2$ ($m_1=1m_0$,$m_2=2m_0$).}
\end{figure*}

\begin{figure}
\centering
\includegraphics[width=0.45\textwidth]{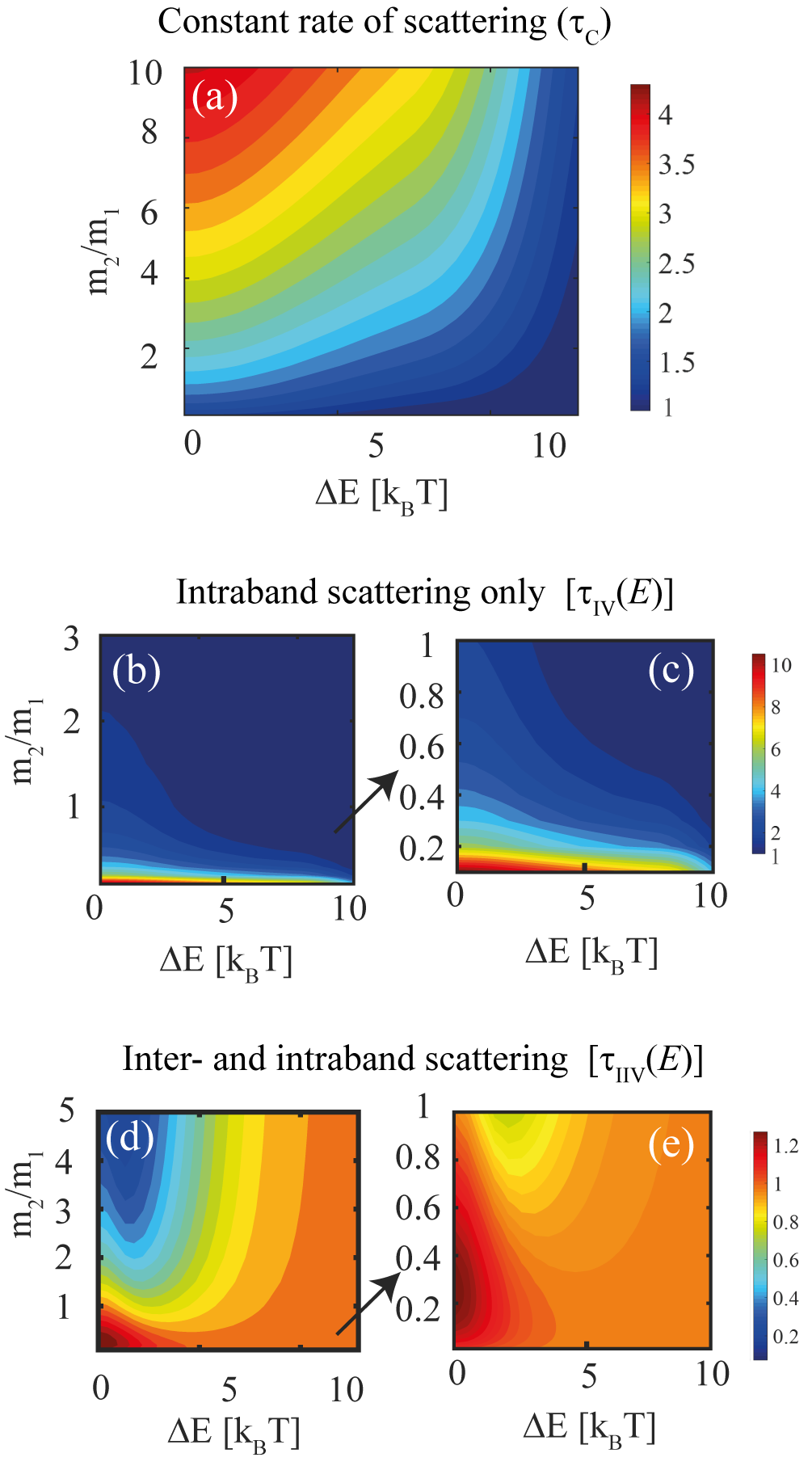}
\caption{\label{fig3}The color plot shows the maximum power factor (PF) for a two-band system of different combinations of band masses. The power factor is normalized by the maximum PF of the case where the two bands are separated by a large energy interval $\Delta E=10\thinspace k_\mathrm{B}T$ (essentially the single band case). Results as a function of the mass ratio $m_1/m_2$ (y-axis) and band separation $\Delta E$ (x-axis) are shown. (a) Constant rate of scattering ( $\tau_\mathrm{c}$) is assumed in the calculation, (b) intra-band scattering only ($\tau_\mathrm{IV}(E)$) with (c) a zoom of (b);  and (d) inter- and intra-band scattering ($\tau_\mathrm{IIV}(E)$) with (e) a zoom of (d). }
\end{figure}

\section{\label{sec:level3}	Band alignment under the parabolic band approximation }

\subsection{\label{sec:level15}Constant scattering rate and time ($\bm{\tau}_\mathbf{C}$) }
Under the constant RTA, $\tau_i(E)$ is a constant and therefore the TD function relation given by Eq.\ref{Eq6} can be simplified to: 

\begin{equation}\label{Eq7}
\Xi\left(E\right)\propto \sum_{i} m_i^{\frac{1}{2}}\ E^\frac{3}{2}\ H(-\Delta E_i).\ \ 
\end{equation}

This indicates that aligning a band of any mass will increase the TD function, resulting finally in an increased conductivity, with the larger masses resulting in larger improvements, a common scenario seen in most band alignment literature. The TD functions for these cases are shown in Figs.\ref{fig1}(c) and \ref{fig1}(d) when bringing in a light/heavy band, respectively. Indeed, it is clear that for the fully aligned cases (blue lines), the TD function is larger when a heavier band is aligned. The thermoelectric coefficients, conductivities, Seebeck coefficients, and power factors, calculated for each scattering scenario, are shown in Fig.\ref{fig2}. Here, the three different panels (1-3) show the results for the three different scattering rate scenarios, $\tau_\mathrm{C}$, $\tau_\mathrm{IV}(E)$, and $\tau_\mathrm{IIV}(E)$, respectively. Within each panel, the left column shows the thermoelectric coefficients $\sigma$, $S$, and PF when a lighter band is aligned, whereas the right column when a heavier band is aligned. Focusing on the left panel, Fig.\ref{fig2}.1, which deals with the TE coefficients results under $\tau_\mathrm{C}$, we see that as a result of the band convergence, the conductivity increases following the increase in the TD function (Fig.\ref{fig2}.1(a, d)). The impact of band alignment on the Seebeck coefficient, however, is not significant (Fig.\ref{fig2}.1(b, e)). The magnitude and sign of the Seebeck coefficient are related to the asymmetry of the electron transport around the Fermi level\cite{corps2015interplay,roychowdhury2017enhanced}, which is indicated by the energy gradient of the TD function. For the range of band effective masses we are concerned with, aligning does not additionally introduce significant asymmetry (or significant change in gradient of the TD function) in the electron transport window around the Fermi level where the PF peak is observed. The resulting power factor shows an improvement upon band alignment (a maximum improvement of 143\% when a band of heavier mass is brought to the band edge, as opposed to 69\% when a band with lighter mass is aligned), as seen in Fig.\ref{fig2}.1(c, f). Therefore, in the case of a constant relaxation time ($\tau_\mathrm{C}$), which is the most commonly employed approximation in theoretical investigations, a power factor improvement is always achieved upon band alignment, with a heavier second band being preferred. As we show further below, this is not the case when $\tau =\tau(E)$. 

\subsection{\label{sec:level12}Intra-band scattering only ($\bm{\tau}_{\mathbf{IV}}(\bm{E})$)}

Within Fermi's Golden Rule, the scattering rate is proportional to the density of available states that a charge carrier can scatter into\cite{lundstrom2009fundamentals,rudan2015physics,qiu2015first}. Therefore, it is natural to investigate the effect of such a scattering scenario on the TE coefficients under band alignment. In multi-valley multi-band  materials the selection rules for each scattering mechanism dictate if the carriers are allowed to scatter only within their current band (intra-band) in the current valley (intra-valley), or whether scattering into states in other bands (inter-band) and other valleys (inter-valley) is also allowed\cite{ lundstrom2009fundamentals, witkoske2017thermoelectric, pshenay2010effect, pshenay2018effect}. For intra-band scattering (limited to intra-valley in multi-valley materials), $\tau_\mathrm{IV}(E)$, we have $\tau_i(E)\propto1/\mathrm{DOS}_i(E)$, and the TD function given by Eq.\ref{Eq6} can be simplified to: 

\begin{equation}
    \Xi(E)\propto \sum_{i}\frac{\ E}{m_i\ }H\left(-\Delta E_i\right)
\end{equation}

This indicates again that aligning a band of any mass will increase the TD function, resulting in an increased conductivity. As opposed to the previous $\tau_\mathrm{C}$ scenario, however, lighter bands benefit transport (since the band mass is now in the denominator), and bringing lighter bands closer to the band edge will provide a higher improvement. This is indicated in Figs.\ref{fig1}(e) and \ref{fig1}(f) for the TD functions in this scattering scenario, where the fully aligned blue line  shows largest improvement under light band alignment, in contrast to the $\tau_\mathrm{C}$ case in Fig.1(c) and 1(d). Notice that the TD functions in Fig.\ref{fig1}(e) and \ref{fig1}(f) are non-smooth, highlighting the additional transport component that is added from the second band. The thermoelectric coefficients resulting out of these TD functions as a function of the Fermi level position are shown in the middle panel of Fig.\ref{fig2}. It can be seen that as a result of the band convergence, the conductivity increases (Fig.\ref{fig2}.2(a, d)), the Seebeck coefficient is still not affected noticeably (Figs.\ref{fig2}.2(b), \ref{fig2}.2(e)), and the resulting power factor shows a large improvement of 200\% when a light band is brought in, as opposed to 49\% when a heavier band mass is aligned (Fig.\ref{fig2}.2(c, f)), respectively). Therefore, in the case of $\tau_\mathrm{IV}(E)$ scattering with $\tau_i(E)\propto1/\mathrm{DOS}_i(E)$, a power factor improvement is always achieved upon band alignment, but now a lighter second band is preferred. Note that in this case the TE coefficients are given in arbitrary units, as we do not consider any specific value for the scattering rate other than its energy dependence.

\subsection{Inter- and intra-band scattering ($\bm{\tau}_{\mathbf{IIV}}(\bm{E})$)}

 In this case, carriers are allowed to scatter elastically to the total density of states available at the energy under consideration, without any selection rules, i.e. both inter- and intra-band (with inter- and intra-valley in multi-valley materials) scattering is allowed ($\tau_i(E)\propto1/\sum_{i}{\mathrm{DOS}_i(E)}$). The TD function relation given by Eq.\ref{Eq6} in this case can be simplified to:
 
 \begin{equation} \label{Eq9}
    \Xi\left(E\right)\propto \frac{\sum_{i} m_i^{\frac{1}{2}} E^\frac{3}{2} H\left(-\Delta E_i\right)}{\sum_{i} m_i^ {\frac{3}{2}} E^\frac{1}{2} H\left(-\Delta E_i\right)}
\end{equation}

From Eq.(\ref{Eq9}), since the denominator has a higher mass exponent, it can be deduced that upon full band alignment, the TD function will only increase  when a light band (B2) is brought close to the band edge and is aligned with a heavier  band (B1), i.e. $m_1>m_2$ (see details in Appendix \ref{A1}). The TD functions for this scattering scenario under bringing in a light/heavy mass are shown in Fig.\ref{fig1}(c) and \ref{fig1}(f), respectively. Here also we assume that $\Delta E_i$ is large enough so that initially, before aligning, only B1 contributes to conduction (green lines). When an additional band B2 is gradually brought close to the band edge to be aligned with band B1, three competing effects take place: 1) the presence of the additional conducting states from B2 tends to increase the TD function, 2) the same states increase the scattering out of B1, which tends to reduce the TD function 3) scattering from B2 reduces since there are less states to scatter into in B1 at energies closer to VB$_0$, increasing the TD function. These interdependences do not allow for the significant improvements in the TD, the conductivity, and the PF that were observed in the previous two scattering scenarios. When $m_1<m_2$, at the energy where the second band is reached, the TD experiences a sharp drop due to increased scattering (Fig.\ref{fig1}(h)). This drop is not very notable when $m_1>m_2$  (Fig.\ref{fig1}(g)). The thermoelectric coefficients for this scenario are shown in the right panel of Fig.\ref{fig2}, for cases where we bring in a lighter band (Fig.\ref{fig2}.3(a-c)), and a heavier band (Fig.\ref{fig2}.3(d-f)). In the first case where $m_1>m_2$, the conductivity is improved upon band convergence, even though not as much as observed in the previous scattering scenarios. The power factor shows an improvement of only 26\% upon full alignment. When a heavier mass is aligned, on the other hand in Fig.\ref{fig2}.3(d-f), the conductivity is reduced and as a result, the power factor is reduced (by 58\%) due to increased scattering, as opposed to all the previous cases (compare the blue fully aligned with the green unaligned lines in Fig.\ref{fig2}.3(f)). This indicates that aligning bands is not always advantageous. It is worth mentioning that this reduction is calculated between the $\Delta E=10\thinspace k_\mathrm{B}T$ and $\Delta E=0$ cases. However, when comparing the power factors in a narrower energy region between the $\Delta E=2k_\mathrm{B}T$ and $\Delta E=0$ (compare the blue with the red lines in Fig.\ref{fig2}.3(f)), there is a small improvement. This indicates that if there is an improvement to the power factor or not, depends on initial band separation, $\Delta E$. As shown in inset of Fig.\ref{fig2}.3(e), this increase is due to an increase in the Seebeck coefficient. The change in gradient of the TD function, as the B2 approaches the vicinity of VB$_0$, caused the Seebeck coefficient to reduce slightly (see inset of Fig.\ref{fig2}.3(f)). Upon full alignment, however, the sharp drop disappears increasing the Seebeck coefficient again. This increase in Seebeck coefficient close to the PF peak overcomes the reduction in conductivity upon full alignment, in this situation (inset of Fig.\ref{fig2}.3(d)). Note that such non-monotonic behaviour can also be present in the conductivity, with increased number of carriers and reduced scattering from B2 overcoming the disadvantage of increased scattering from B1.

In the case of 3 bands, where two bands of masses $m_{2}$ and $m_{3}$ are aligned with a band of mass $m_1$, the condition for an improved 3 band TD function when those three bands are completely aligned compared to the single band TD function is given by (see details in Appendix \ref{A3}):

 \begin{equation}
 m_1>\frac{\left(m_2^{3/2}+m_3^{3/2}\right)}{(m_2^{1/2}+m_3^{1/2})}
 \end{equation}
 
 From the above equation, we see that in general terms, in the case of $\tau_\mathrm{IIV}(E)$, bringing in lighter bands into transport is beneficial for the TD function. Note, however, that this trend is not monotonic when considering the power factor versus $\Delta E$ as we will be discussing below.
 
 In order to have a more comprehensive first-order understanding of the benefits of band alignment, we have calculated the power factor using the effective mass approximation for combinations of different band effective mass ratios ($m_2/m_1$) from 0.1 up to 10, and for band separations ($\Delta E$) of up to $10\thinspace k_\mathrm{B}T$, as first adopted by \textit{Jeong et al.}\cite{jeong2011electronic}. The maximum power factors for all cases are shown in the color-plots of Fig.\ref{fig3} for the three scattering scenarios, $\tau_\mathrm{C}$, $\tau_\mathrm{IV}(E)$ and $\tau_\mathrm{IIV}(E)$. For the ranges we have considered, under a constant rate of scattering($\tau_\mathrm{C}$) in Fig.\ref{fig3}(a), aligning a band with any mass is going to improve the power factor (brighter colors towards the left for smaller $\Delta{E})$ and aligning heavier bands is more beneficial (brighter colors towards the top of Fig.\ref{fig3}(a)). This is a result of more states being involved in transport, without however increasing scattering rates, which are kept constant. When energy dependent  $\tau_\mathrm{IV}(E)$  is considered, the transport in each valley and each band is independent of the other and therefore we observe in Fig.\ref{fig3}(b) (and its’ zoomed version Fig.\ref{fig3}(c)), that aligning a band with any mass is going to improve the power factor. However, in this case, it is the lighter bands ($m_2/m_1<1 $) that are more beneficial.
 
 When $\tau_\mathrm{IIV}(E)$ is considered, interestingly, we no longer have a monotonic relationship between the alignment $\Delta E$, mass ratio, and the power factor improvement. When initially the $\Delta E$  is large, i.e. when only one band is initially contributing to conduction within the range we consider ($10\thinspace k_\mathrm{B}T$), aligning masses heavier than the existing mass ($m_2/m_1>1$) is going to reduce the power factor, as shown in Fig.\ref{fig3}(d).  This shows that, counter-intuitively, aligning bands is not always beneficial for the power factor. This is because the benefit of increase in conduction band states is offset by increase in scattering. For $m_2/m_1>1$, when $\Delta E$  is gradually reduced, the peak power factor is reduced at first and reaches a minimum, but then it experiences an increase (still less than the max power factor at band separation $\Delta E=10\thinspace k_\mathrm{B}T$). One reason for this non-monotonic behavior is an increase in the Seebeck coefficient closer to VB$_0$ where the power factor peak is observed as a result of asymmetry in the electron transport introduced by the placement of the second band. Another reason for this non-monotonic behavior is the increment seen in conductivity as a result of competing effects of increased conduction states and scattering, as $\Delta E$ becomes smaller (contribution from B2 keeps increasing with reducing $\Delta E$ as a result of reduced number states in B1 to scatter into, closer to VB$_0$). Therefore, for certain initial smaller $\Delta E$, even under the $m_2/m_1>1$ condition, we will observe an improvement in the power factor with both reducing and increasing $\Delta E$ due to non-monotonic behavior in the conductivity and the Seebeck coefficient. Similar non-monotonic behavior is seen when lighter bands are aligned ($m_2/m_1>1$), but unlike for heavier bands, improvements to the power factor can be seen under most $m_2/m_1$ ratios (areas with colors close to red seen in Fig.\ref{fig3}(e)), even when the initial $\Delta E$  is large. Based on Fig.\ref{fig3}(d), note that it is possible to improve the power factor though misaligning the bands (increasing $\Delta E$) for certain $m_2/m_1$ ratios.
 
 There is an optimum $m_2/m_1$ ratio which gives the best power factor under $m_2/m_1<1$ as seen in Fig.\ref{fig3}(e). To provide an indication about the band mass ratio for the maximum PF, we find the $m_2/m_1$ that gives the maximum TD function, when bands are fully aligned. The TD function for 2 bands can be written as:
 
 \begin{equation}
 \Xi\left(E\right)\propto\mathrm{\ \ (}\frac{1\ +p^\frac{1}{2}\ }{1+p^\frac{3}{2}\ })\frac{E}{m_1}
 \end{equation}
 \\
 where $p=m_2/m_1$ (Appendix \ref{A2}). By taking the derivative of  $\Xi(E)$ with respect to the mass ratio $p$, we can find the ratio that maximizes the $\Xi(E)$. We find this value to be $p=0.25$ and this corresponds to the value of $m_2/m_1$ that gives the maximum power factor in Figs.\ref{fig3}(e). 
 
 with the highest improvement seen when the bands are completely aligned ($\Delta E=0$) under $\tau_\mathrm{C}$ and  $\tau_\mathrm{IV}(E)$ scattering scenarios. For the mass ranges we have considered, under $\tau_\mathrm{C}$, aligning heavy masses are more beneficial, but under $\tau_\mathrm{IV}(E)$, aligning light bands are more beneficial. Under $\tau_\mathrm{IIV}(E)$ the outcome is more complex and whether there is an improvement or not depends on masses of the bands and the initial $\Delta E$. In general, for $\tau_\mathrm{IIV}(E)$, again aligning lighter masses is more beneficial under most initial $\Delta E$ values. 

\section{\label{sec:level4}Non-parabolic band (NPB) approximation results for Co-based half-Heuslers}

\begin{figure}[hbt!]
\centering
\includegraphics[width=0.45\textwidth]{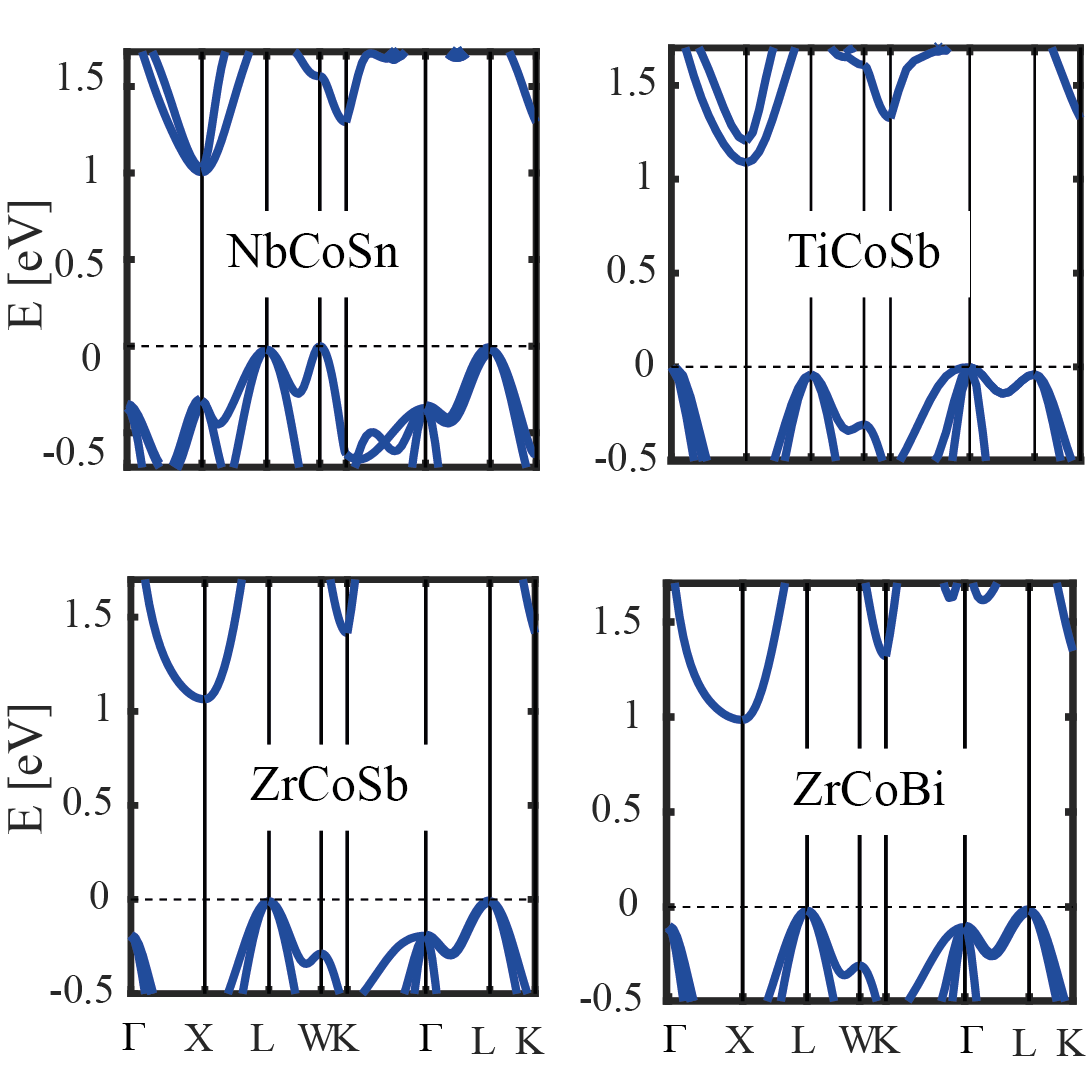}
\caption{\label{fig4} Bandstructures of the half-Heuslers NbCoSn, TiCoSb, ZrCoSb, and ZrCoBi.}
\end{figure}

\begin{figure*}[hbt!]
\centering
\includegraphics[width=1\textwidth]{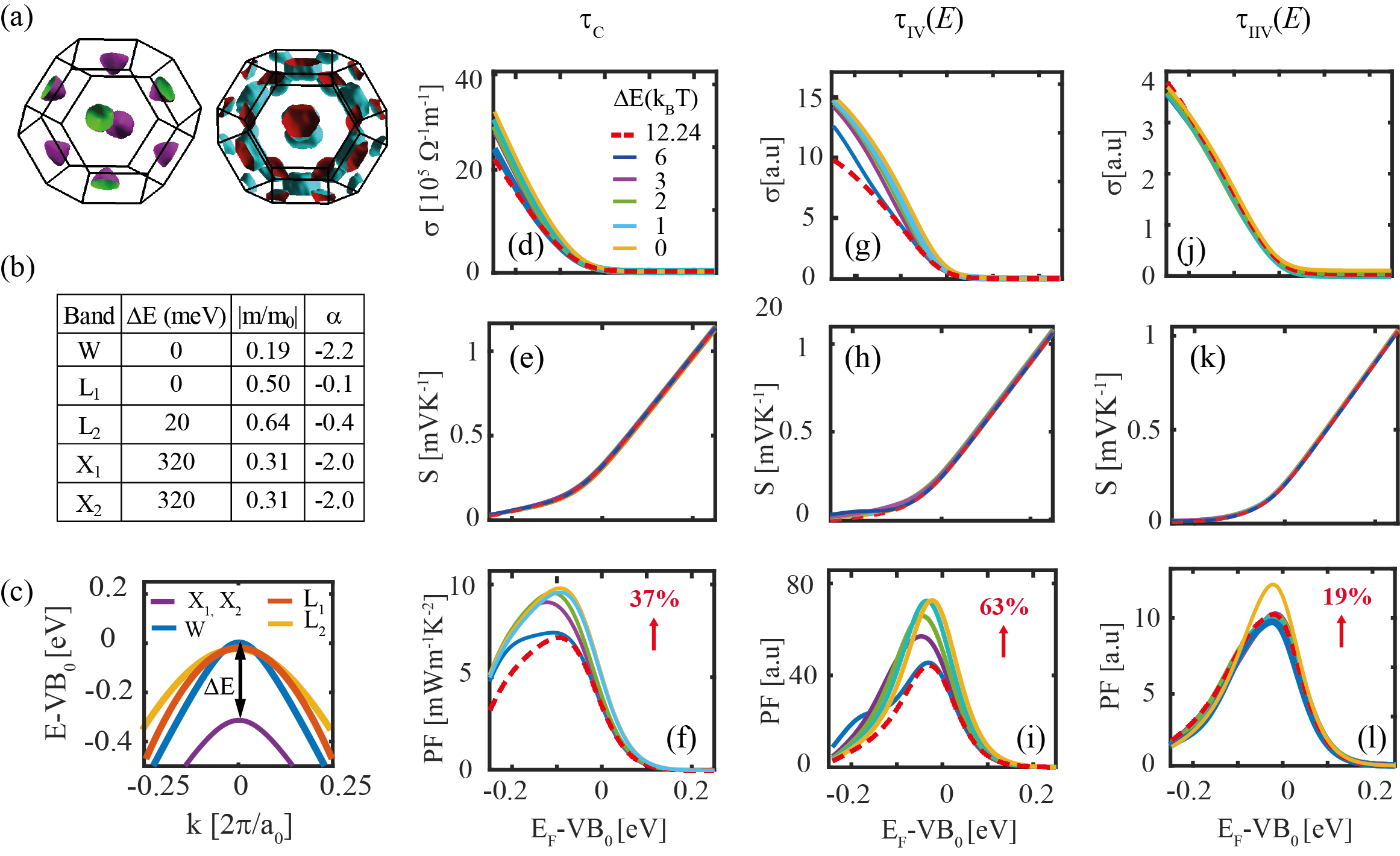}
\caption{\label{fig5} Thermoelectric coefficients for band alignment in NbCoSb described using the non-parabolic band (NPB) approximation, under three different scattering scenarios. (a)  The Fermi surface of NbCoSn at 0.1~eV below VB$_0$. (b) NPB parameters calculated for valence bands at W, L (L$_1$ and L$_2$) and X (X$_1$ and X$_2$) valleys, assuming transport in the [100] direction. (c) The resulting NbCoSb bands extracted using the NPB approximation. The energy position of bands X$_1$ and X$_2$ is shifted by $\Delta E$ (in units of $k_\mathrm{B}T$) until it is fully aligned with the VB$_0$. Column-wise: (d)-(f) Thermoelectric coefficients ($\sigma$, $S$, and PF) calculated using the NPB approximation for NbCoSn under constant rate of scattering ($\tau_\mathrm{C}$). (g-i) $\sigma$, $S$ and PF under intra-band\textbackslash intra-valley scattering only ($\tau_\mathrm{IV}(E)$). (j-l) $\sigma$, $S$ and PF under inter- and intra-band\textbackslash inter- and intra-valley scattering ($\tau_\mathrm{IIV}(E)$). The percentage improvement given is the peak to peak improvement between the $\Delta E=0$ and  $\Delta E=12.24k_\mathrm{B}T$  (given by the yellow-solid line and red-dashed lines respectively).}
\end{figure*}

\begin{figure*}[hbt!]
\begin{minipage}[c]{0.75\textwidth}
\centering
\includegraphics[width=0.98\textwidth]{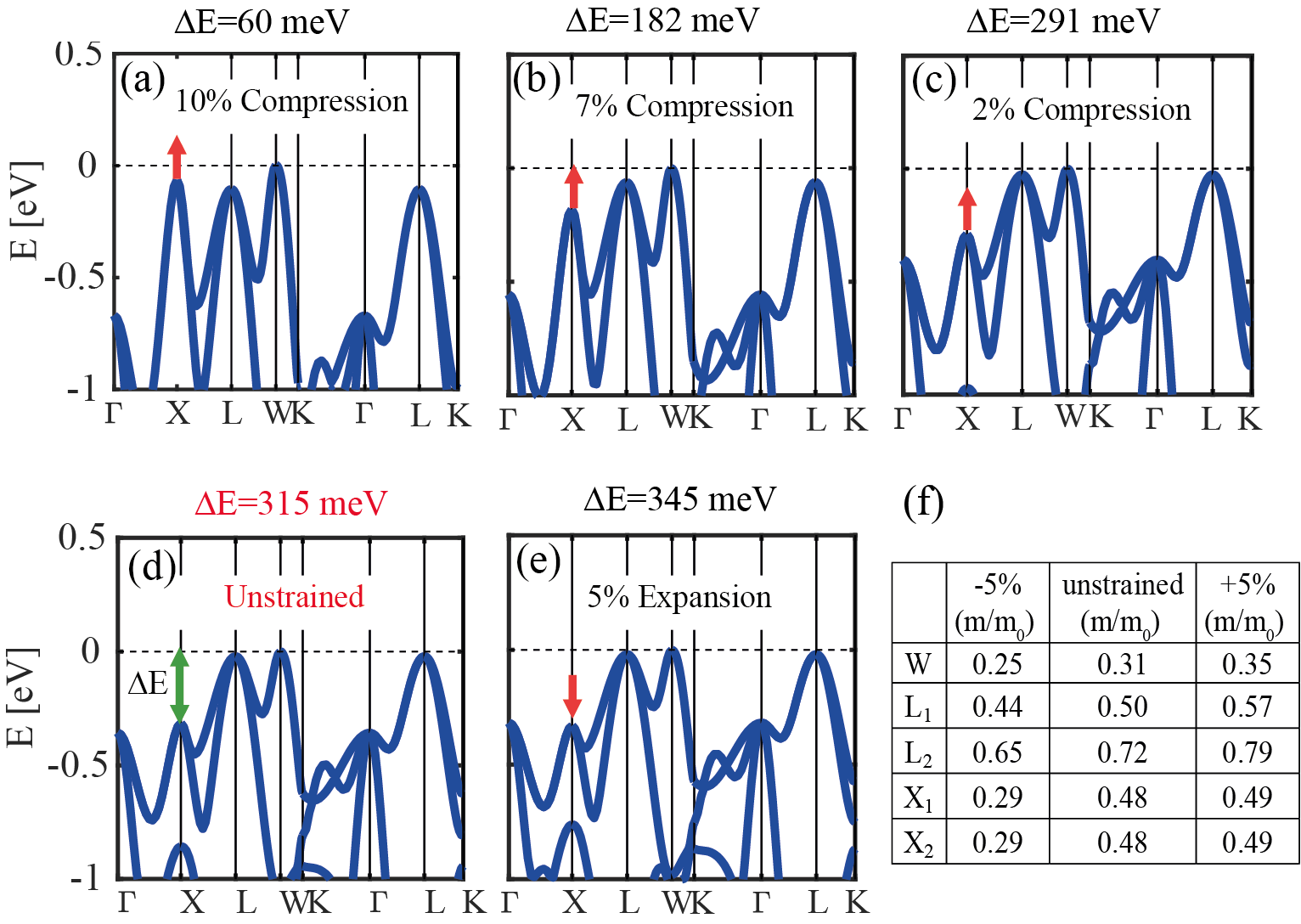}
\end{minipage}\hfill
\begin{minipage}[c]{0.22\textwidth}
\caption{\label{fig6} (a)-(e) Application of strain in NbCoSb to align the X valley with the valence band edge. Strain values are as indicated in the figures, with (d) showing the unstrained bandstructure. (f) The calculated effective masses for the relevant bands at L,W and X valleys of NbCoSn using the parabolic band approximation for different strain levels [-5\% (compressive), unstrained, and +5\% (tensile)]. The energy separation $\Delta E$ between the X valley and VB$_0$ are noted above the sub-figures.}
\end{minipage}
\end{figure*}

\begin{figure}
\centering
\includegraphics[width=0.45\textwidth]{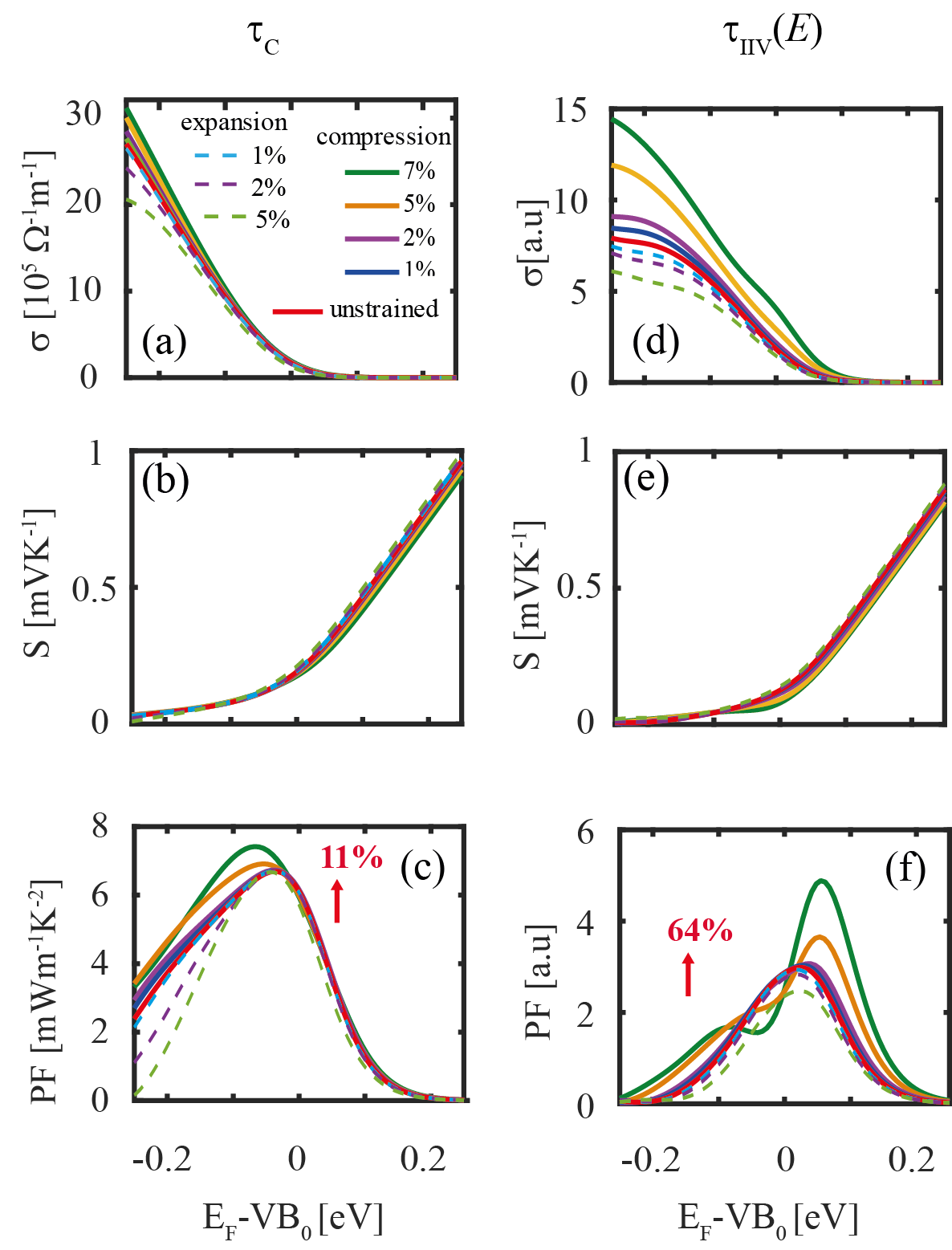}
\caption{\label{fig7} Column-wise: (a)-(c) Thermoelectric coefficients ($\sigma$, $S$, and PF) calculated using the non-parabolic band (NPB) approximation for NbCoSn under constant rate of scattering ( $\tau_\mathrm{C}$). (d)-(f) $\sigma$, $S$, and PF under inter- and intra-band \textbackslash inter- and intra-valley scattering ($\tau_{IIV}(E)$). The percentage improvement given is the peak to peak improvement between the unstrained ($\Delta E=315$~meV) and 7\% compressive strain  $(\Delta E=182$~meV$)$, given by the red-solid lines and green-solid lines, respectively.}
\end{figure}

\begin{figure}
\centering
\includegraphics[width=0.45\textwidth]{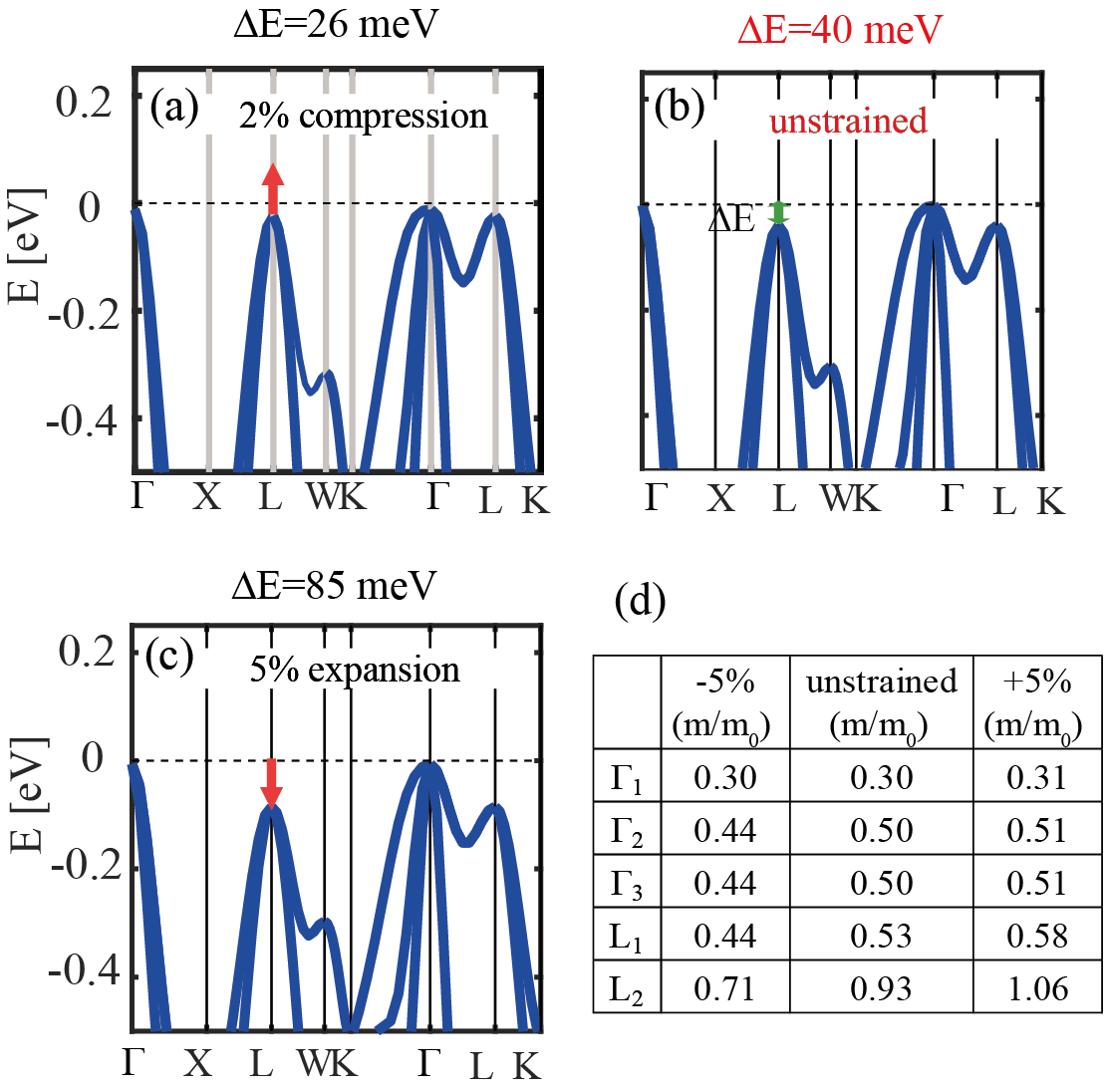}
\caption{\label{fig8}(a)-(c) Application of strain in TiCoSb to align the L valley with the valence band edge. Strain values are as indicated in the figures, with (b) showing the unstrained bandstructure. (d) The calculated effective masses for the relevant bands at L and $\Gamma$ valleys of TiCoSb using the parabolic band approximation for different strain levels [-5\% (compressive), unstrained, and +5\% (tensile)]. The energy separation $\Delta E$ between the L valley and VB$_0$ are noted above the sub-figures.}
\end{figure}

\begin{figure}
\centering
\includegraphics[width=0.45\textwidth]{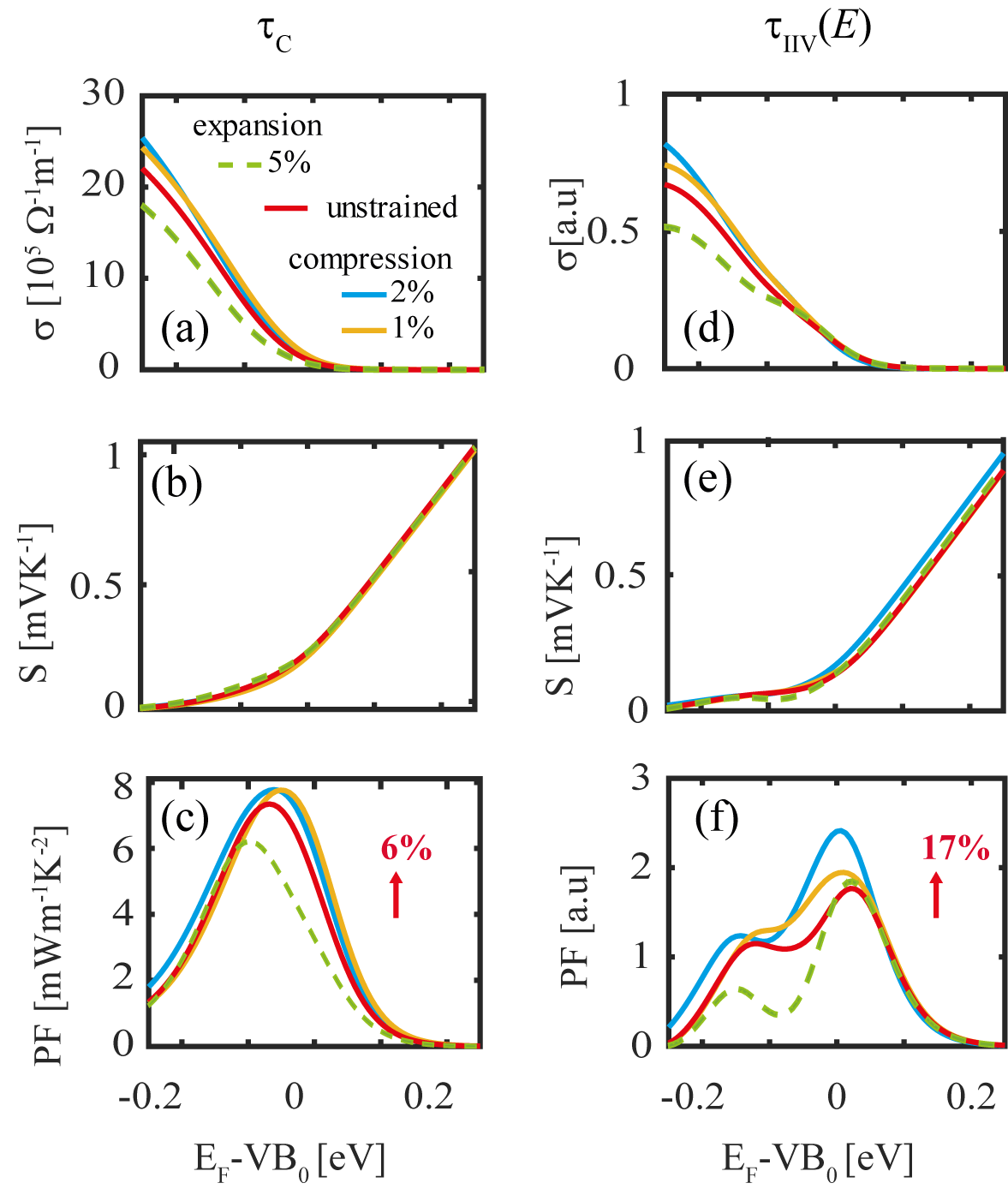}
\caption{\label{fig9}Column-wise: (a)-(c) Thermoelectric coefficients ($\sigma$ , $S$ , and PF) calculated using the non-parabolic band (NPB) approximation for TiCoSb under constant rate of scattering ( $\tau_\mathrm{C}$). (d)-(f) $\sigma$,$S$ , and PF under inter- and intra-band\textbackslash inter- and intra-valley scattering ($\tau_{IIV}(E)$). The percentage improvement given is the peak to peak improvement between the unstrained ($\Delta E=40$~meV) and 2\% compressive strain  ($\Delta E=26$~meV), given by the red-solid lines and blue-solid lines, respectively.}
\end{figure}

Here onwards, we start our investigations of band alignment in Co-based half-Heuslers. When examining the DFT extracted bandstructures of the four half-Heuslers, NbCoSn, TiCoSb, ZrCoSb and ZrCoBi shown in Fig.\ref{fig4}, it is apparent that multiple bands from several valleys are available close to the VB$_0$. For instance, in NbCoSn in Fig.\ref{fig4}(a), in addition to the bands at the L and W points that are already aligned at the VB$_0$, there exist heavy and light bands at the X and $\Gamma$ points within 0.3\thinspace eV of the VB$_0$. Aligning these bands that are in the vicinity of the VB$_0$, particularly the bands at the X point that have a large equivalent valley degeneracy of 3, can lead to an improved conductivity and power factor. Similar features that are useful for bandstructure engineering strategies to be applied can be seen in the other three materials as well, especially near the VB$_0$. The TE coefficients for these materials for n-type and \textit{p}-type cases versus the Fermi level position, extracted numerically using BoltzTraP, under a constant relaxation time, $\tau_\mathrm{C}$, are shown in Appendix \ref{A3}. From here on, we have selected to use band alignment to further improve the \textit{p}-type power factor, however, similar studies could be performed for improving the n-type material power factor as well. 

As a first approach in examining how these materials would behave if certain bands are brought closer, or completely aligned with the VB$_0$, we pick two of the materials in Fig.\ref{fig4}, namely NbCoSn (because of high initial $\Delta E$ between the X and L valleys) and TiCoSb (of a lower initial $\Delta E$ between $\Gamma$ and L). We then used the simple non-parabolic band approximation to create the essential features of the bandstructures of these materials in the [100] transport direction. Once we form and calibrate our approximate bands to the DFT bands, we can control their alignment at will, without worrying at the moment about how this alignment will be achieved in practice. The $E-\mathbf{k}$ relation for a non-parabolic band is given by:

\begin{equation}
    E\left(1+\alpha E\right)=\frac{\hbar^2\mathbf{k}^{2}}{2m}\ 
\end{equation}

where  $\alpha$ is the non-parabolicity parameter. The parameters DOS$_i(E)$ and $v_i(E)$ that are necessary to calculate the TD function using Eq.(6) can be also analytically calculated under the NPB approximation for 3D bands as:

\begin{equation}
    \textrm{DOS}_i(E)=\frac{m_i^{\frac{3}{2}}}{\pi^2\hbar^3}N_i\sqrt{2E\left(1+\alpha_iE\right)}\left(1+2\alpha_iE\right)
\end{equation}
\begin{equation}
    v_i(E)=\ \sqrt{\frac{2E\left(1+\alpha_iE\right)-\Delta E_i}{m_i}} \frac{1\ }{\left(1+2\alpha_iE\right)}
\end{equation}

The Fermi surface of NbCoSn at an energy 0.1\thinspace eV below the VB$_0$ is shown in Fig.\ref{fig5}(a). This captures the two bands seen at L and W points of NbCoSn (see bandstructure in Fig.\ref{fig4}). Despite the fact that the bands are strongly curved, we find that a NPB approximation can fit the bands reasonably well, at least up to 0.25\thinspace eV below VB$_0$. The bandstructure contains two already aligned bands at the L point (L$_1$ and L$_2$) and one band at W (W$_1$) with equivalent valley degeneracies of 4 and 6, respectively (Fig.\ref{fig4}). The two bands at the X  (X$_1$ and X$_2$), with equivalent valley degeneracies of 3 are positioned at 315\thinspace meV ($12.24k_\mathrm{B}T$ at $T=300$\thinspace K) below VB$_0$, and these are the bands that we will align with VB$_0$. The non-parabolic model parameters that describe the relevant bands at  X, L and W valleys are given in the table of Fig.\ref{fig5}(b). Here we assume the [100] direction as the transport direction. Figure \ref{fig5}(c) shows the bands reconstructed using the NPB approximation. To verify the accuracy of the NPB approximation using the parameters that we have extracted for NbCoSn, in Fig.\ref{fig5}(b) we have compared the power factor of the bandstructure we constructed with that of the DFT bandstructure using a fully numerical calculation performed using BoltzTraP in Appendix \ref{A5}. Good agreement is found between the NPB and DFT bandstructures for $E_F$ positioned up to -0.1\thinspace eV (4\thinspace$k_\mathrm{B}T$ into the valence band), which is anyway beyond where the $E_F$ needs to be placed at for maximum PF, or under realistic scenarios. Beyond $E=-0.15$\thinspace eV the two methods slightly diverge, signaling that the shape of the actual bandstructure cannot be mapped to a NBP approximation beyond those energies. 

After constructing an equivalent simplified bandstructure, we now proceed in extracting the electrical conductivity, Seebeck coefficient and PF for NbCoSn under gradual alignment of bands X$_1$ and X$_2$. We extract the TE coefficients under the three different scattering cases we described above ($\tau_\mathrm{C}$- Fig.\ref{fig5}(d-f), $\tau_\mathrm{IV}(E)$- Fig.\ref{fig5}(g-i), $\tau_\mathrm{IIV}(E)$- Fig.\ref{fig5}(j-l)). The alignment energy step we impose is in units of $k_\mathrm{B}T$ as indicated in the caption of Fig.\ref{fig5}(d). With the dashed-red line we show the results for the original NPB bandstructure, before attempting any alignment. Upon band alignment, improvements to the conductivity, and hence to the power factor, can be observed in all three scattering scenarios. No significant variations appears in the Seebeck coefficient. The highest improvement is achieved when the bands are fully aligned (yellow lines, for $\Delta E =0$). Since the bands we are aligning are effectively lighter than two of the bands that are already aligned at VB$_0$, the most improvement (63\% in Fig.\ref{fig5}(i)) is seen under $\tau_\mathrm{IV}(E)$ scattering, because states with higher velocities are brought into transport, as discussed earlier. Smaller improvements are achieved in the other two scattering cases, the $\tau_\mathrm{C}(E)$ and the $\tau_\mathrm{IIV}(E)$ (37\% in Fig.\ref{fig5}(f) and 19\% in Fig.\ref{fig5}(l), respectively), as the advantage of bringing in high velocity states can be utilized in all cases, as explained above. 

The corresponding band alignment results using the NPB approximation for the second material we consider, namely TiCoSb, is shown in Appendix 6. In summary, improveents to the power factor are seen under all three scattering scenarios for this material as well, with the most improvement (39\% in Fig.\ref{fig15}(l)) seen under ${\tau}_\mathrm{IIV}({E})$).

\section{\label{sec:level5}The influence of realistic alignment through strain on the power factor}
\subsection{\label{sec:level6}The use of strain to achieve band alignmenet}

In reality, band alignment can be achieved using a variety of methods such as applying strain \cite{capellini2014tensile, jeong2014design}, alloying \cite{pei2011convergence,low2012electronic,zhou2016lead}, increasing temperature (as in skutterudites \cite{tang2015convergence} and lead tellurides \cite{lalonde2011lead}), etc. Here, for the purposes of our investigation into the influence of band alignment on the PF, we use the easier method within DFT, which is the use of hydrostatic strain, either compressive or tensile. We investigate the effect of strain in the bandstructures of three of the half-Heuslers we consider, NbCoSn, TiCoSb and ZrCoSb. In the following sections, for each material, the thermoelectric coefficients were calculated from DFT derived bandstructures numerically using BoltzTraP (under the constant relaxation time, $\tau_\mathrm{C}$, approximation) and our own codes (for $\tau_\mathrm{IIV}(E)$) still by using the DFT extracted numerical DOS and velocities (i.e. we do not use either the parabolic, or the non-parabolic band approximations in this section). Due to difficulty in obtaining valley specific velocities and density of states from DFT, in this section we do not consider the $\tau_\mathrm{IV}(E)$ scattering case. We use these examples to highlight the different PF observations under different band conditions that can take place. Because our purpose is to provide an indication as to what alignment will do to the PF, and not how alignment can in practice be achieved, in our study we sometimes use strain even up to unrealistic values (i.e. 10\% in some cases), until close to full alignment is achieved, for example. Large distortions in lattices could be achieved by alloying, however, due to the larger computational complexity, we perform the calculations using strain.    

\subsection{\label{sec:level7}NbCoSn under strain}
\begin{figure*}
\begin{minipage}[c]{0.75\textwidth}
\centering
\includegraphics[width=1\textwidth]{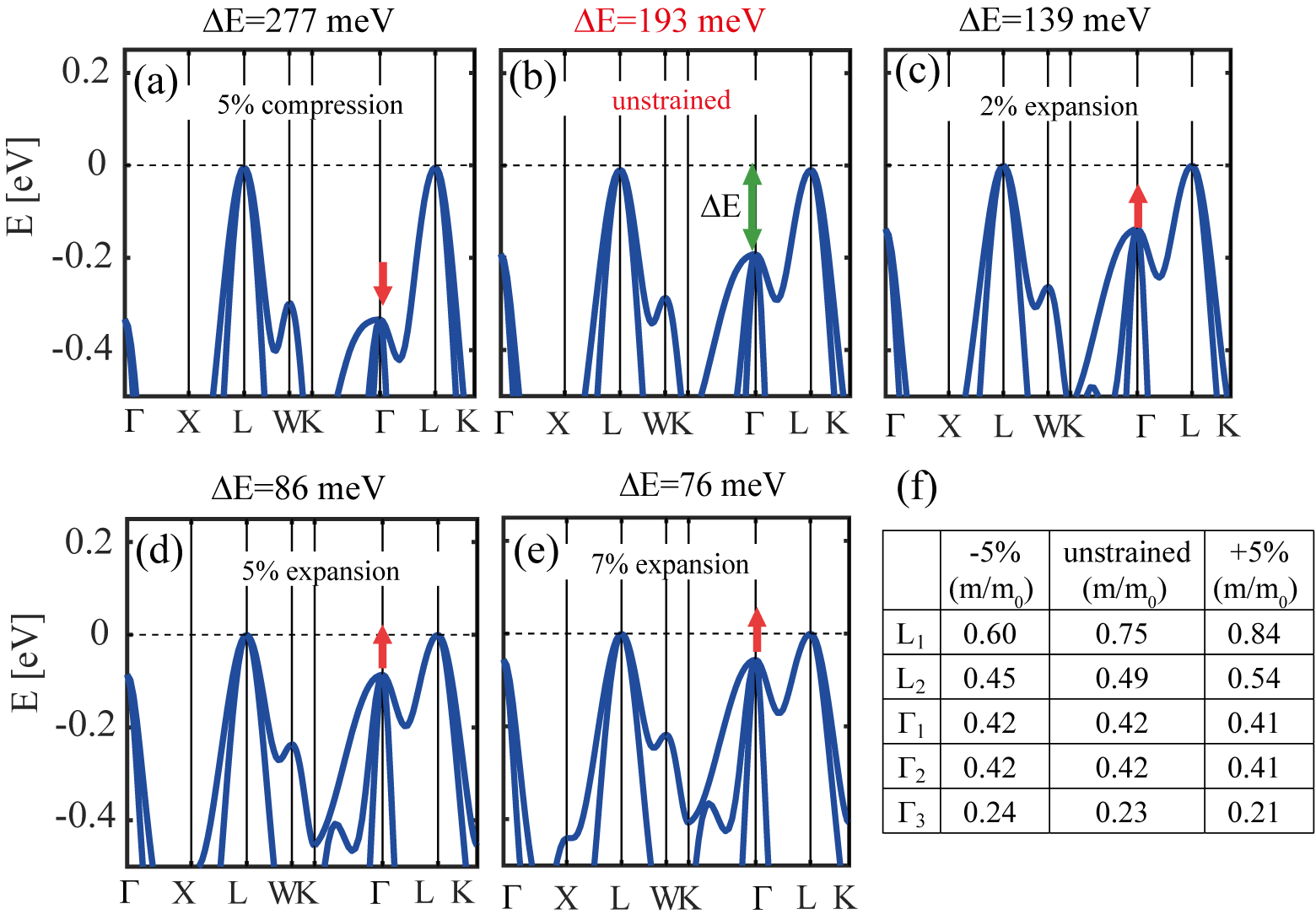}
\end{minipage}\hfill
\begin{minipage}[c]{0.22\textwidth}
\caption{\label{fig10}(a)-(e) Application of strain in ZrCoSb to align the $\Gamma$ valley with the valence band edge. Strain values are as indicated in the figures, with (b) showing the unstrained bandstructure. (f) The calculated effective masses for the relevant bands at L and $\Gamma$ valleys of ZrCoSb using the parabolic band approximation for different strain levels [-5\% (compressive), unstrained, and +5\% (tensile)]. The energy separation $\Delta E$ between the $\Gamma$ valley and VB$_0$ are noted above the sub-figures.  }
\end{minipage}
\end{figure*}

\begin{figure}[h]
\centering
\includegraphics[width=0.45\textwidth]{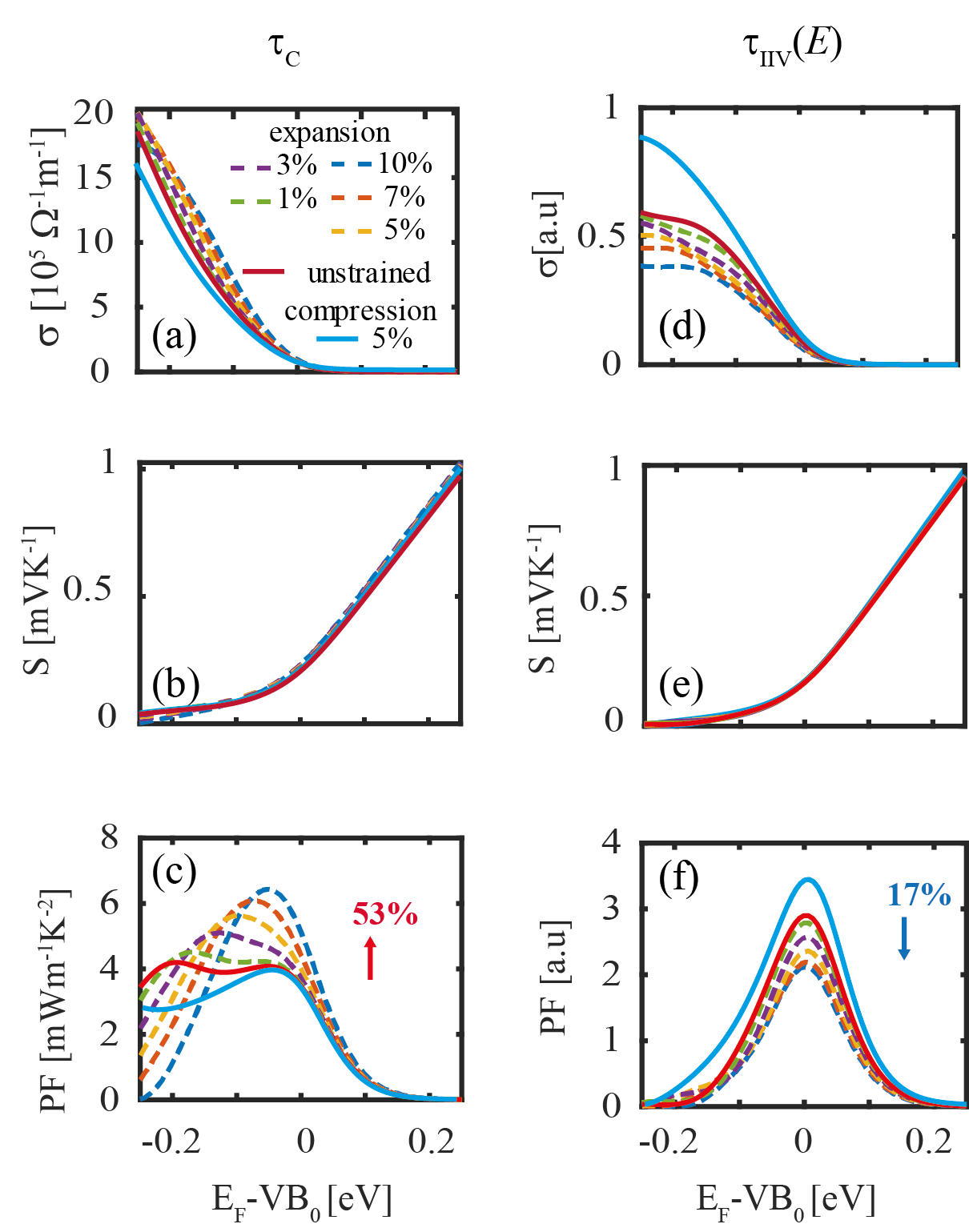}
\caption{\label{fig11} Column-wise: (a)-(c) Thermoelectric coefficients ($\sigma$, $S$, and PF) calculated using the non-parabolic band approximation for ZrCoSb under constant rate of scattering ( $\tau_\mathrm{C}$). (d)-(f) $\sigma$, $S$, and PF under inter- and intra-band\textbackslash inter- and intra-valley scattering ($\tau_\mathrm{IIV}(E)$). The percentage improvement given is the peak to peak improvement between the unstrained ($\Delta E=193$~meV) and 10\% tensile strain  ($\Delta E=27$~meV), given by the red-solid lines and blue-dashed lines, respectively.}
\end{figure}

The bands of NbCoSn can be manipulated with compression and expansion as shown in Figure \ref{fig6}(a-e). When compressed, the bands at the X point are brought closer to VB$_0$, reducing $\Delta E$. To fully align, a compressive strain as large as 10\% is required. When the material is under tensile strain (expansion), $\Delta E$ increases (see red arrows in Fig.\ref{fig6}(a-e) for the energy shift of the bands in each strain case). It is important to note, however, that the curvatures of the bands also change with strain. The effective masses reduce with compressive strain and increase with tensile strain as documented in the table of Fig.\ref{fig6}(f). Figure \ref{fig7} shows the TE coefficients conductivity, Seebeck coefficient and PF for the two scattering cases considered ($\tau_\mathrm{C}$ -Fig.\ref{fig7} first column, and $\tau_\mathrm{IIV}(E)$- Fig.\ref{fig7} second column). The PF is improved in both scattering scenarios under compressive strain, which aligns the bands (solid lines). The improvement, however, is larger in the case of $\tau_\mathrm{IIV}(E)$ compared to constant scattering $\tau_\mathrm{C}$, something that was not observed in the simple NPB analysis we performed earlier in Section IV. The fact that the masses reduce with band convergence is unfavorable under a constant scattering rate. It is, however, favorable under $\tau_\mathrm{IIV}(E)$ as it brings bands with higher velocities that results in less scattering within the transport window, in the cases where $\Delta E$ is large. Note here the difference between the simplified NPB approximation extracted TE coefficients as shown earlier in Fig.\ref{fig5}. In that case, the constant relaxation time approximation provided larger benefits to the power factor. However, under the realistic alignment scenario, where additional effects such as the masses reduction with band convergence appear, the $\tau_\mathrm{IIV}(E)$ scattering scenario gives a better improvement.

\subsection{\label{sec:level8}TiCoSb under strain}
The next material we attempt to manipulate with strain is TiCoSb. The energy separation $\Delta E$ between bands at L and $\Gamma$  points of $\approx 40$\thinspace meV can be reduced by applying compressive strain, leading to band convergence, as shown in Fig.\ref{fig8}(a-c). The curvatures of all the bands close to VB$_0$ are reduced with compressive strain and is increased with tensile strain as documented in the Table in Fig.\ref{fig8}(d). Figure \ref{fig9} shows the TE coefficients conductivity, Seebeck coefficient and PF for the two scattering cases considered ($\tau_\mathrm{C}$ - Fig.\ref{fig9} first column and $\tau_\mathrm{IIV}(E)$– Fig.\ref{fig9} second column). Under a constant rate of scattering, we only see 6\% improvement when the bands are fully aligned using compressive strain (Fig.\ref{fig9}(c)), but the improvement is 17\% under $\tau_\mathrm{IIV}(E)$ scattering (Fig.\ref{fig9}(f)). The fact that masses reduce with alignment as before is favorable under the latter scattering scenario, which has contributed to the larger improvement observed.  The $\Delta E$ value in this situation is only 40\thinspace meV ($1.55\thinspace k_\mathrm{B}T$ at $T=300$\thinspace K), i.e. bands are almost aligned even without strain. The bands can be fully aligned by applying only $\approx 2\%$ compressive strain (a much more realistic value compared to the ones needed for NbCoSn). Another observation, different compared to NbCoSn is that the bands we are aligning have higher masses than the bands that are already aligned. Although this is favorable under $\tau_\mathrm{C}$, only moderate improvements to the conductivity, and hence to the power factor, are observed in this case, because $\Delta E$ is only 40\thinspace meV to begin with, and the masses reduce with band alignment. Again, the highest improvement is seen when the bands are fully aligned. The higher improvement that is observed in the PF under $\tau_\mathrm{IIV}(E)$, is benefitted by the fact that bands become lighter as they are aligned. This behavior was seen earlier in Fig.\ref{fig3}(d). With expansion, which misaligns the bands (increases $\Delta E$), under the $\tau_\mathrm{IIV}(E)$ scenario, the conductivity and therefore the power factor, increases slightly as seen in Figs.\ref{fig9}(d) and \ref{fig9}(f) (compare the green-dotted line with the solid red line), in contrast to $\tau_\mathrm{C}$ case. Similar non-monotonic behavior was observed in Fig.\ref{fig3}(d). The percentage improvement values differ from the NPB approximation calculations because of the change in the masses.

\subsection{\label{sec:level9}ZrCoSb under strain}
The third material we examine under strain is ZrCoSb (Fig.\ref{fig10}). The bands at the $\Gamma$ point ($\Gamma_1$,$\Gamma_2$ and $\Gamma_3$) reside 193\thinspace meV (or $7.48\thinspace k_\mathrm{B}T$ with $T=300$\thinspace K) below VB$_0$ (Fig.\ref{fig10}(b)). We attempt to align them with the bands at the L point (L$_1$ and L$_2$) at VB$_0$. As opposed to the previous two materials, band convergence can be achieved by applying tensile strain (expansion), rather than compression (Fig.\ref{fig10}(a-e)). The already aligned bands L$_1$ and L$_2$ become lighter with compressive strain and heavier with tensile strain, as documented in the Table of Fig.\ref{fig10}(f). Increase in band masses is beneficial under a constant scattering time $\tau_\mathrm{C}$, as explain above in Section III. Therefore, in the thermoelectric coefficients in Fig.\ref{fig11} ($\tau_\mathrm{C}$- first column – Fig.\ref{fig11}(a-c)), we find a substantial improvement in the power factor when the bands are fully aligned under $\tau_\mathrm{C}$ ($\approx$53\% in Fig.\ref{fig11}(c) for 7\% strain, but $\approx$20\% for a more realistic strain value of 3\%). However, this increase in the effective masses and alignment of heavier masses are unfavorable in general under the $\tau_\mathrm{IIV}(E)$ (second column – Fig.\ref{fig11}(d-f)) scattering scenario. Therefore, we see a decrease in the power factor by 17\% when the bands are aligned under $\tau_\mathrm{IIV}(E)$ in Fig.\ref{fig11}(f) (compare the red-solid line to the blue-dashed line). In the case of compressive strain, which further misaligns the $\Gamma$ valley, but reduces the effective mass of the already aligned bands at L valley, we see an increase in the power factor (red-solid line vs blue-solid line), as lighter masses are favorable under the $\tau_\mathrm{IIV}(E)$ scenario, as explained in Section III. In the case of 5\% compression in Fig.\ref{fig11}(f), therefore, we observe a 12\% improvement.

\begin{figure}[htb!]
\centering
\includegraphics[width=0.45\textwidth]{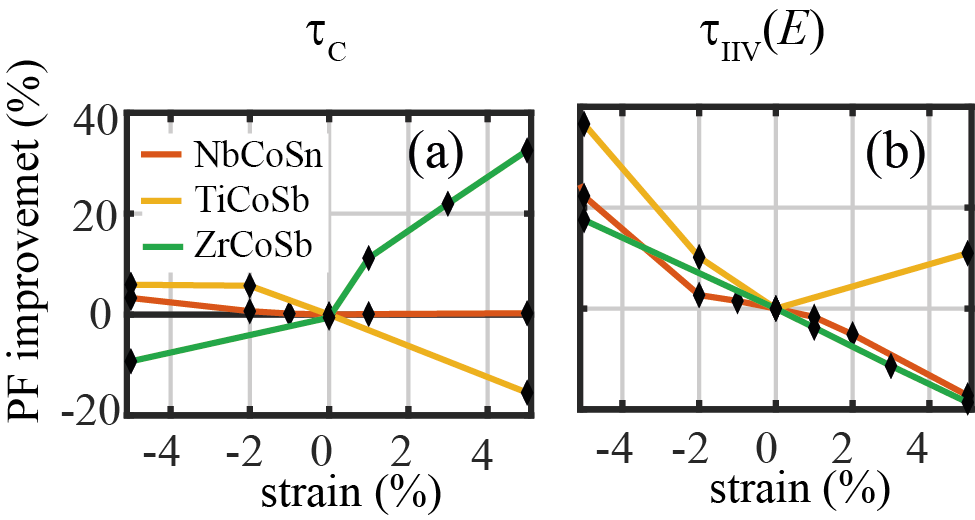}
\caption{\label{fig12}Percentage improvement of the power factor (from peak to peak) under: (a) constant rate of scattering ( $\tau_\mathrm{C}$), and (b) inter- and intra-band\textbackslash inter- and intra-valley scattering ($\tau_\mathrm{IIV}(E)$)
versus strain percentage is applied for NbCoSn, TiCoSb and ZrCoSb. Positive strain indicates expansion while negative strain indicates compression. }
\end{figure} 

The strain required to completely align the bands can be high as $\approx$10\% in NbCoSn and ZrCoSb, which is unrealistic to achieve. However, as seen in Figs.\ref{fig7}, \ref{fig9} and \ref{fig11}, reducing the distance $\Delta E$ is sufficient to see an improvement to the PF, even though fully aligning the bands gives the maximum improvement. Our aim, however, beyond indicating the possibilities of band alignment with strain, was also to demonstrate different scenarios of how the bands' mass can change with alignment, and that this is a factor that can change the expectations out of simple models and needs to be taken into consideration. Importantly, the changes can be different for different materials and strain conditions. In Fig.\ref{fig12}, finally, we show a summary of the power factor improvements with strain of only up to 5\% for the three materials we examined (note that we did not perform a strain study for the ZrCoBi since the bandstructure looks similar to the other half-Heuslers we examined). In Fig.\ref{fig12}(a), under a constant scattering rate,  ZrCoSb (green lines in Fig.\ref{fig12}) shows the best performance improvement with tensile strain. Here, a $\approx 15\%$ improvement in the power factor can be seen even with 1-2\% tensile strain, but the improvement jumps to $\approx 30\%$  at 5\% strain. TiCoSb shows nearly 5\% improvement when 1-2\% compressive strain is applied. The outcomes, however, are different under $\tau_\mathrm{IIV}(E)$ scattering. It is compressive strain that allows for PF improvements for all three materials under band alignment up to values between 20\%-40\% at 5\% strain, which is quite significant. Tensile strain degrades the performance of ZrCoSb and NbCoSn, but still allows improvements for TiCoSb. Interestingly, in the case of $\tau_\mathrm{IIV}(E)$ scattering, TiCoSb allows PF improvements with either compressive or tensile strain. However, to achieve these benefits in practice, the precise scattering conditions need to be identified, as different conditions lead to different conclusions.

\section{Conclusion}

In this work, we have provided a comprehensive investigation into the benefits of band alignment (or band convergence) of complex bandstructure thermoelectric materials in improving the power factor, by going beyond the constant relaxation time approximation. After a generic investigation using the alignment of simple parabolic bands, we used the actual DFT extracted bandstrucstures of the \textit{p}-type Co-based half-Heuslers NbCoSn, TiCoSb and ZrCoSb, and used strain to align the various bands that appear in the valence band. Using the Boltzmann transport equation under the relaxation time approximation, we explored the band alignment effect on the power factor under three different scattering conditions (as the detail scattering physics of half-Heuslers are still not known): i) constant relaxation time approximation – as is common in most literature, ii) scattering rates proportional to the density of final states, but under only intra-band\textbackslash intra-valley considerations, and iii) scattering rates proportional to the density of final states, with both inter- and intra-band\textbackslash inter- and intra-valley scattering considerations. We showed that the outcome of band alignment can be completely different in each of the different scattering cases. Specifically, constant relaxation time scenarios favour alignment of heavier bands (i.e. bringing heavier bands closer to lighter ones) for larger improvements, as those provide more transport states, but without increase in scattering. On the other hand, alignment of lighter bands is favoured for the second scattering situation, where only intra-band (and intra-valley) scattering is considered, as they provide a small number of carriers with higher velocities, and do not interfere significantly with the scattering of carriers in already aligned bands. The third scattering scenario where both intra- and inter-band (with inter- and intra-valley) scattering is considered, shows a more complicated non-monotonic relationship between power factor benefits, band separation and mass ratio. Under this scattering scenario, band convergence can lead to reduction of the power factor in certain cases particularly when heavy bands are aligned, due to increased carrier scattering offsetting the advantage of increase in conducting states. Because of this, we showed that it is the misalignment of the bands, instead of the alignment, which leads to power factor improvements in certain situations. We show that there is an optimum band separation - mass ratio combination, to obtain the best PF improvements.  In general, however, we showed that aligning lighter bands favours the power factor in this situation as well. We stressed that aiming for a multi-band multi-valley bandstructure does not always improve thermoelectric performance and band alignment strategies need to consider the scattering physics, as they determine whether the power factor will increase, or decrease upon alignment. In addition, we point out that under band alignment in realistic material engineering scenarios (i.e. by applying strain), the band curvature can be changed, which adds another complication in determining whether alignment can help or not. In general, however, under constant scattering rate scenarios, bringing in heavy masses helps the power factor, whereas under density of states dependent rates, bringing in lighter masses is what helps. With regards to the application of strain to improve the power factor of half-Heuslers, we showed that application of strain up to 5\% can improve the power factor by up to 40\%, but whether this is achieved by compression or tension, depends on the specific material and which valleys are aligned. Thus, our work stresses the importance of more accurate theoretical treatment for each material in examining its thermoelectric properties, as the specific details can lead to different conclusions. 


\begin{acknowledgments}
Acknowledgements: This work has received funding from the European Research Council (ERC) under the European Union’s Horizon 2020 Research and Innovation Programme (Grant Agreement No. 678763). 
\end{acknowledgments}

\appendix
\section{\label{A1}	Condition for TD function improvement under $\tau_\mathrm{IIV}(E)$ scattering: Case of 2 bands}

When  $\Delta E$ is large, only the first band B1 contributes to conduction. Therefore, from Eq.(\ref{Eq9}) the TD function can be written as:
\begin{equation}
    \Xi\left(E\right)\propto\frac{m_1^{\frac{1}{2}}\ E^\frac{3}{2\ }}{m_1^{\frac{3}{2}}\ E^\frac{1}{2\ }\ } \tag{A1}
\end{equation}

When bands B1 and B2  are completely aligned, i.e. $\Delta E=0$, both B1 and B2 contribute to conduction. Therefore, from Eq.(\ref{Eq9}), the TD function can be written as:
\begin{equation}
\Xi\left(E\right)\propto\frac{m_1^{\frac{1}{2}}\ E^\frac{3}{2\ }+m_2^{\frac{1}{2}}\ E^\frac{3}{2\ }}{m_1^{\frac{3}{2}}\ E^\frac{1}{2\ }+m_2^{\frac{3}{2}}\ E^\frac{1}{2\ }}\tag{A2}
\end{equation}

To have gains by alignment, we set the TD function given by Eq.(A2) to be larger than what is given by Eq.(A1), which leads to:
\begin{equation}
\frac{m_1^{\frac{1}{2}}\ E^\frac{3}{2\ }+m_2^{\frac{1}{2}}\ E^\frac{3}{2\ }}{m_1^{\frac{3}{2}}\ E^\frac{1}{2\ }+m_2^{\frac{3}{2}}\ E^\frac{1}{2\ }}>\frac{m_1^{\frac{1}{2}}\ E^\frac{3}{2\ }}{m_1^{\frac{3}{2}}\ E^\frac{1}{2\ }\ }\tag{A3}
\end{equation}

\begin{equation}
\Rightarrow \left(m_1^{\frac{1}{2}}\ +m_2^{\frac{1}{2}}\right)m_1>\ m_1^{\frac{3}{2}}\ +m_2^{\frac{3}{2}}\tag{A4}
\end{equation}

\begin{equation}
\Rightarrow m_1>m_2\tag{A5}
\end{equation}

\section{\label{A2}	Band masses that maximize the TD function under $\tau_\mathrm{IIV}(E)$: Case of 2 bands}

When the bands B1 and B2 are completely aligned,  $\Delta E=0$, and both B1 and B2 contribute to conduction.  We set  $m_2/m_1\ =p$, and from Eq.(\ref{Eq9}) the TD function can be written as:
\begin{equation}
\Xi\left(E\right)\propto\frac{m_1^{\frac{1}{2}}\ E^\frac{3}{2\ }+p^\frac{1}{2}m_1^{\frac{1}{2}}\ E^\frac{3}{2\ }}{m_1^{\frac{3}{2}}\ E^\frac{1}{2\ }+{p^\frac{3}{2}m}_1^{\frac{3}{2}}\ E^\frac{1}{2\ }}\tag{B1}
\end{equation}

\begin{equation}
\Rightarrow \Xi\left(E\right)\propto\left(\frac{1+p^\frac{1}{2}}{1+p^\frac{3}{2}\ }\ \right)\frac{E}{m_1}\tag{B2}
\end{equation}
Therefore, the value of the mass ratio $p$ that maximizes the TD function can be obtained by:
\begin{equation}
\frac{d}{dp}\Big(\frac{1\ +p^\frac{1}{2}\ }{1+p^\frac{3}{2}\ }\Big)=0  \tag{B3}
\end{equation}

\begin{equation}
\Rightarrow \frac{3\left(p^\frac{1}{2}+1\right)p-(1+p^\frac{3}{2})}{(1+p^\frac{3}{2})}=0 \tag{B4}
\end{equation}

\begin{equation}
\Rightarrow 2p^\frac{3}{2}+3p-1=0 \tag{B5}
\end{equation}

The solution to Eq.(B5) gives $p=0.25$.

\section{\label{A3}	Condition for TD function improvement under $\tau_\mathrm{IIV}(E)$  scattering: Case of 3 bands }

When the separation of bands, $\Delta E$, is large, only the first band B1 contributes to conduction. Therefore, from Eq.(\ref{Eq9}) the TD function can be written as:

\begin{equation}
    \Xi\left(E\right)\propto\frac{m_1^{\frac{1}{2}}\ E^\frac{3}{2\ }}{m_1^{\frac{3}{2}}\ E^\frac{1}{2\ }\ }\tag{C1}
\end{equation}
When all three bands B1, B2 and B3 are completely aligned, $\Delta E=0$, and all B1, B2 and B3 bands contribute to conduction. Therefore, from Eq.(\ref{Eq9}) the TD function can be written as:

\begin{equation}
\Xi\left(E\right)\propto\frac{m_1^{\frac{1}{2}}\ E^\frac{3}{2\ }+m_2^{\frac{1}{2}}\ E^\frac{3}{2}+m_3^{\frac{1}{2}}\ E^\frac{3}{2\ }}{m_1^{\frac{3}{2}}\ E^\frac{1}{2\ }+m_2^{\frac{3}{2}}\ E^\frac{1}{2\ }+m_3^{\frac{3}{2}}\ E^\frac{1}{2\ }}\tag{C2}
\end{equation}

By setting the TD function given by Eq.(C2) to be larger than what is given by Eq.(C1), we find the condition of the mass of B1 compared to that of B2 and B3 for TD improvement as:
\begin{equation}
\frac{m_1^{\frac{1}{2}}\ E^\frac{3}{2\ }+m_2^{\frac{1}{2}}\ E^\frac{3}{2}+m_3^{\frac{1}{2}}\ E^\frac{3}{2\ }}{m_1^{\frac{3}{2}}\ E^\frac{1}{2\ }+m_2^{\frac{3}{2}}\ E^\frac{1}{2\ }+m_3^{\frac{3}{2}}\ E^\frac{1}{2\ }}\ \ >\frac{m_1^{\frac{1}{2}}\ E^\frac{3}{2\ }}{m_1^{\frac{3}{2}}\ E^\frac{1}{2\ }\ }\tag{C3}
\end{equation}

\begin{equation}
\Rightarrow m_1\ >\ \frac{m_2^{\frac{3}{2}}+m_3^{\frac{3}{2}}}{m_2^{\frac{1}{2}}\ +m_3^{\frac{1}{2}}}\tag{C4}
\end{equation}

\section{\label{A4}Calculation of transport coefficients of half-Heuslers}

\begin{figure}[!htb]
\centering
\includegraphics[width=0.3\textwidth]{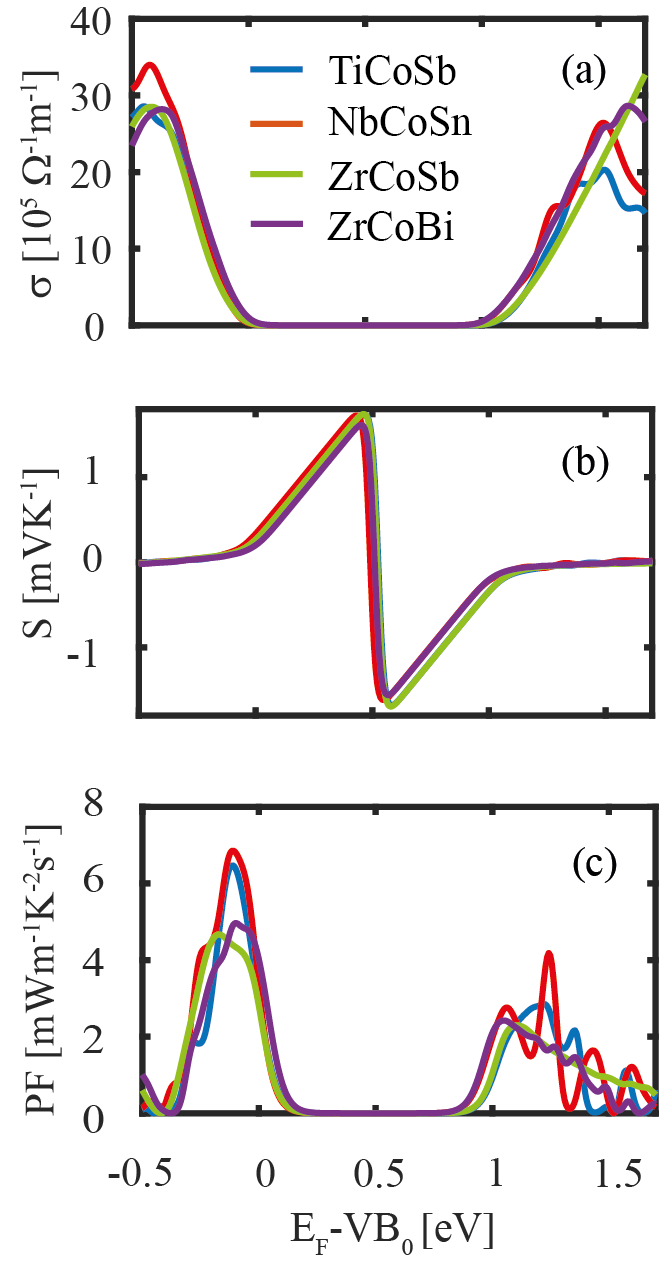}
\caption{\label{fig13}(a) Conductivity, (b) Seebeck coefficient and the (c) power factor of NbCoSn, TiCoSb, ZrCoSb and ZrCoBi for both electrons and holes, calculated using BoltzTraP, employing the constant relaxation time approximation with $\tau={10}^{-14}$\thinspace s.}
\end{figure}
 Figure \ref{fig13} compares the conductivity, Seebeck coefficient and the power factor for the half-Heuslers NbCoSn, TiCoSb, ZrCoSb and ZrCoBi, under a constant relaxation time of $10^{-14}$\thinspace s. They all have similar power factors, but TiCoSb and NbCoSn are slightly better when the \textit{p}-type power factor is considered.

\section{\label{A5}Comparison of the non-parabolic band (NPB) approximation with the full-band calculations}

\begin{figure}[h]
\centering
\includegraphics[width=0.45\textwidth]{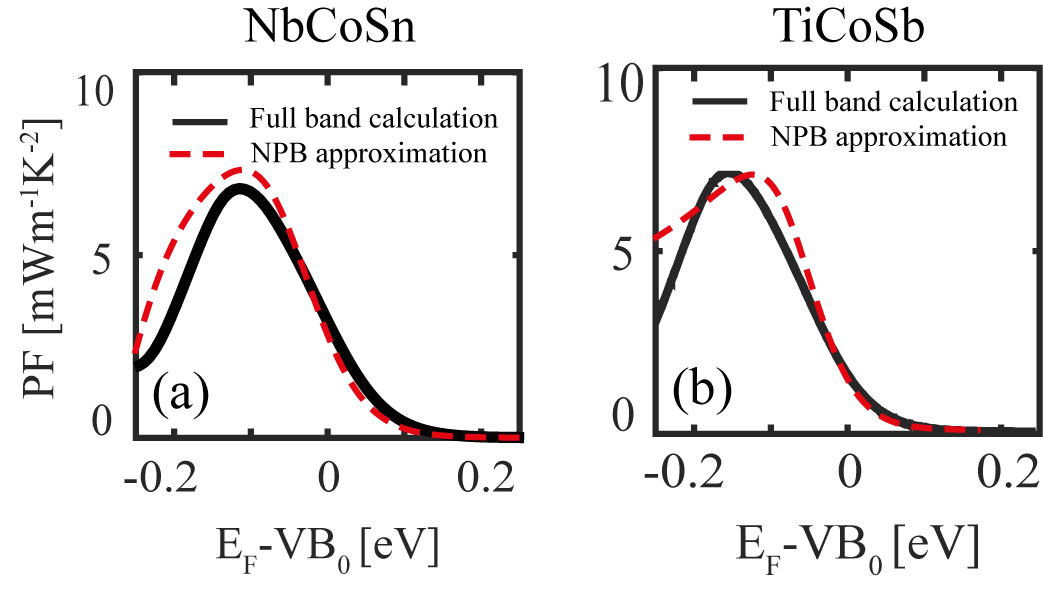}
\caption{\label{fig14}Comparison of results from the non-parabolic band (NPB) approximation(red-dashed lines) and numercal full-band calculations (black-solid lines) using BoltzTraP, under the constant relaxation time approximation with $\tau={10}^{-14}$\thinspace s, for (a) NbCoSn and (b) TiCoSb.}
\end{figure}
The outcome of the NPB approximation is compared with the full-band (DFT derived) results obtained from BoltzTraP calculations, where no approximation about the band shape is made. The NPB parameters used for Figs.\ref{fig14}(a) and (b) are given in Figs.\ref{fig5}(b) and \ref{fig15}(b), respectively. The left and right figures show the comparison for NbCoSn and TiCoSb, respectively. Black-solid lines are power factor results calculated using the full band structure with BoltzTraP while the red-dashed lines are calculated using the NPB approximation and our Boltzmann transport codes. A good match is observed between the two models, indicating the validity of the NPB approximation for the relevant energies under consideration, when the NPB parameters used are extracted from DFT calculated bands.

\section{\label{A6} Thermoelectric coefficients for TiCoSb under the NPB approximation}
\begin{figure*}[t]
\centering
\includegraphics[width=0.95\textwidth]{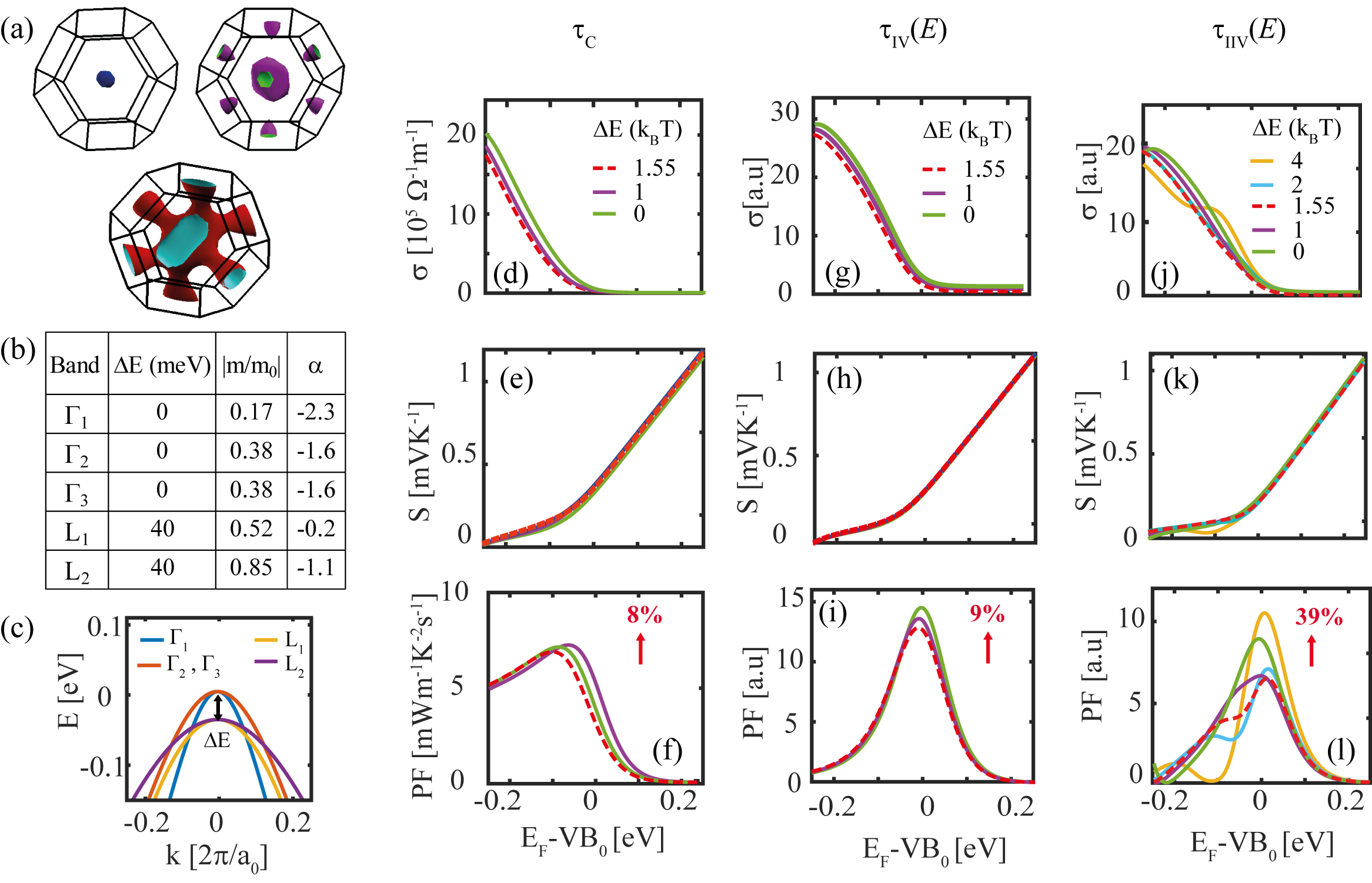}
\caption{\label{fig15}Thermoelectric coefficients for band alignment in TiCoSb described using the non-parabolic band (NPB) approximation, under three different scattering scenarios. (a) The Fermi surface of TiCoSb at 0.1\thinspace eV below VB$_0$. (b) NPB parameters calculated for valence bands at $\Gamma$ ($\Gamma_{1}, \Gamma_{2}$ and $\Gamma_{3}$) and L (L$_1$ and L$_2$) valleys, assuming transport in the [100] direction. (c) The resulting TiCoSb bands extracted using the NPB approximation. The energy position of bands L$_1$ and L$_2$ are shifted (in units of $k_\mathrm{B}T$) until it is fully aligned with the VB$_0$). Column-wise: (d)-(f) Thermoelectric coefficients ($\sigma$, $S$, and PF) calculated using the NPB approximation for TiCoSb under constant rate of scattering ($\tau_\mathrm{C}$). (g-i) $\sigma$, $S$, and PF under intra-band\textbackslash intra-valley scattering only($\tau_\mathrm{IV}(E)$). (j-l) $\sigma$, $S$, and PF under inter- and intra-band\textbackslash inter- and intra-valley scattering ($\tau_\mathrm{IIV}(E)$). The percentage improvement given is the peak-to-peak improvement between $\Delta E=0$ and $\Delta E=1.55 k_\mathrm{B}T$  (given by the yellow-solid and red-dashed lines respectively).}
\end{figure*}
In this section we examine the possibility of improving the power factor by using band alignment in TiCoSb, as shown in Fig.\ref{fig15}, using non-parabolic bands extracted from the DFT bandstructure. The Fermi surface of TiCoSb at energy 0.1\thinspace eV below the VB$_0$ is shown in Fig.\ref{fig15}(a). The two bands at L point (L$_1$ and L$_2$) are only 40\thinspace meV below VB$_0$, which is not a significant separation. Therefore, we would not expect significant improvements upon aligning those bands. We still perform this analysis, however, to demonstrate that the trends here are different compared to what observed in the main text for NbCoSn, and that in this case one observed in a real material the peculiar non-monotonic trend that is shown in Fig.\ref{fig3}(e). In Fig.\ref{fig15}(b) we present the NPB approximation parameters that describe bands at L point (L$_1$ and L$_2$) and $\Gamma$ point($\Gamma_1$,$\Gamma_2$ and $\Gamma_3$) assuming the transport to be in the [100] direction. Again, to verify the accuracy of the NPB approximation for TiCoSb, we compare in Appendix \ref{A5} the PFs of the bandstructure we constructed (Fig.\ref{fig15}(c)) with a fully numerical calculation done on a DFT derived bandstructure using BoltzTrap and found good agreement between two methods within -0.2eV ($8\thinspace k_\mathrm{B}T$ into the valence band, which is sufficient for our study). Beyond -0.2\thinspace eV, PF calculated for the two methods deviate, since the shape of the actual bandstructure diverges from the NPB shape at higher energies. 

Now we move on to calculate thermoelectric parameters for TiCoSb under the NPB approximation for different scattering scenarios ($\tau_\mathrm{C} $- Fig.\ref{fig15}(d-f), $\tau_\mathrm{IV}(E)$ - Fig.\ref{fig15}(g-i), and $\tau_\mathrm{IIV}(E)$- Fig.\ref{fig15}(j-l)), under different alignment levels in units of $k_\mathrm{B}T$. The results for the original bandstructure, before attempting any alignment is shown by the dashed-red lines. The bands we are aligning have higher masses than the bands already aligned. This is the favorable PF improvement condition under a constant rate of scattering and and moderate improvements of $\approx 8$\% are observed. Moderate improvements to the conductivity and hence to the power factor can also be observed in second scattering scenarios. The improvements are small, however, because $\Delta E=1.55\thinspace k_\mathrm{B}T$, only, and on the other hand, no significant variation is observed in the Seebeck coefficient. The largest improvement (of $\approx 39$\%) is observed when bands are close together, but not fully aligned (yellow lines) under $\tau_\mathrm{IIV}(E)$. This is a combined effect of increase in conductivity by bringing another band together, but without increasing yet scattering from the light band into the heavy band, while also slightly increases the Seebeck coefficient due to the presence of the second band – i.e. the band alignment is fine-tuned as in the situation described in Fig.\ref{fig3}(e).

\bibliography{References.bib}

\end{document}